\documentclass[submission, Phys]{SciPost}

\usepackage[T1]{fontenc}

\usepackage{listings}

\usepackage{fvextra}
\usepackage{epsfig}
\usepackage{amsmath,amssymb}
\usepackage{graphicx}
\usepackage{hyperref}
\hypersetup{colorlinks=true,urlcolor=blue,citecolor=blue,linkcolor=blue,breaklinks=true}
\usepackage[noabbrev]{cleveref}
\usepackage{adjustbox}
\usepackage{floatrow}
\usepackage{footnote}
\usepackage{grffile}
\usepackage{color}
\usepackage{xcolor}
\usepackage{dsfont}
\usepackage{mathtools}
\usepackage{verbatim}
\usepackage{bbold}
\usepackage{hhline}
\usepackage{textcomp}
\usepackage{comment}
\usepackage[section]{placeins}

\usepackage[percent]{overpic}
\usepackage[normalem]{ulem}
\usepackage{diagbox}

\usepackage{physics}
\definecolor{unirot}{RGB}{165,30,55}
\definecolor{unigelb}{RGB}{168,148,99}
\definecolor{pythongreen}{RGB}{0,128,0}
\definecolor{unigrau}{RGB}{50,65,75}

\usepackage[bottom]{footmisc} 
\definecolor{LightGray}{gray}{0.9}
\definecolor{CodeKeyword}{RGB}{0,64,128}
\definecolor{CodeComment}{RGB}{0,110,0}
\definecolor{CodeString}{RGB}{160,40,20}
\definecolor{CodeNumber}{RGB}{90,90,90}

\lstdefinestyle{codeblock}{
   frame=lines,
   framesep=2mm,
   backgroundcolor=\color{LightGray},
   basicstyle=\footnotesize\ttfamily,
   numbers=left,
   numberstyle=\tiny\color{CodeNumber},
   keywordstyle=\color{CodeKeyword}\bfseries,
   commentstyle=\color{CodeComment}\itshape,
   stringstyle=\color{CodeString},
   breaklines=true,
   showstringspaces=false,
   keepspaces=true,
   columns=fullflexible
}

\lstnewenvironment{cppcode}{
   \lstset{style=codeblock,language=C++,tabsize=2}
}{}

\lstnewenvironment{makecode}{
   \lstset{style=codeblock,language=make,tabsize=4}
}{}

\lstnewenvironment{pythoncode}{
   \lstset{style=codeblock,language=Python,tabsize=4}
}{}

\lstnewenvironment{bashcode}{
   \lstset{style=codeblock,language=sh,tabsize=2}
}{}

\usepackage[compat=1.1.0]{tikz-feynman}
\usetikzlibrary{positioning}
\usetikzlibrary{arrows.meta}
\usetikzlibrary{decorations.markings}
\usetikzlibrary{decorations.pathmorphing}
\usetikzlibrary{decorations.pathreplacing}
\usetikzlibrary{calc}
   
\tikzset{
   linePlain/.style={draw=black, thick},
   linePlainBlue/.style={draw=blue, thick},
   lineBare/.style={draw=gray, thick},
   lineWithArrowEnd/.style={draw=black, thick, postaction={decorate},decoration={markings,mark=at position 1. with {\arrow[scale=1.2]{latex}}}},
   lineWithSmallArrowEnds/.style={draw=gray, postaction={decorate},decoration={markings,mark={at position 1. with {\arrow[scale=0.8]{latex}} }}},
   lineBareWithArrowEnd/.style={draw=gray, thick, postaction={decorate},decoration={markings,mark=at position 1. with {\arrow[scale=1.2]{latex}}}},
   lineWithArrowCenter/.style={draw=black, thick, postaction={decorate},decoration={markings,mark=at position .6 with {\arrow[scale=1.2]{latex}}}},
   lineWithArrowCenterCenter/.style={draw=black, thick, postaction={decorate},decoration={markings,mark=at position .5 with {\arrow[scale=1.2]{latex}}}},
   lineBareWithArrowCenter/.style={draw=black, thick, postaction={decorate},decoration={markings,mark=at position .6 with {\arrow[scale=1.2]{latex}}}},
   lineWithArrowCenterEnd/.style={draw=black, thick, postaction={decorate},decoration={markings,mark=at position .85 with {\arrow[scale=1.2]{latex}}}},
   lineWithArrowCenterEnd/.style={draw=black, thick, postaction={decorate},decoration={markings,mark=at position .85 with {\arrow[scale=1.2]{latex}}}},
   lineWithArrowCenterStart/.style={draw=black, thick, postaction={decorate},decoration={markings,mark=at position .35 with {\arrow[scale=1.2]{latex}}}},
   lineWithArrowInline/.style={draw=black, semithick, postaction={decorate},decoration={markings,mark=at position .7 with {\arrow[scale=1.2]{latex}}}},
   vertex/.style={draw, shape=circle, fill=black, minimum size=1.1mm, inner sep=0mm, outer sep=0mm},
   bosonLine/.style={draw=black, thick, decorate, decoration={snake, segment length=2mm, amplitude=0.6mm}},
}

\newcommand{\cpp}[1]{\lstinline[language=C++,basicstyle=\small\ttfamily]!#1!}

\newcommand{\directory}[1]{\path{#1}}

\newcommand{\tikzm}[2]{
   \tikz[baseline=-0.65ex]{#2}
}

\newcommand{\bosonfull}[4]{
   \draw[bosonLine, very thick] (#1,#2) -- (#3,#4);
}

\newcommand{\arrowslefthalffullbis}[3]{
   \def\shift{0.3};
   \def\shiftbox{0.3*#3};
   \coordinate (center) at (#1,#2);
   \coordinate (bottomleft)  at ($(center) + (-\shiftbox,-\shiftbox)$);
   \coordinate (topleft)     at ($(center) + (-\shiftbox,+\shiftbox)$);
   \draw[linePlain] (bottomleft)  -- ($(bottomleft) + (-\shift,-\shift)$);
   \draw[linePlain] ($(topleft)     + (-\shift,+\shift)$) -- (topleft);
}

\newcommand{\arrowsrighthalffullbis}[3]{
   \def\shift{0.3};
   \def\shiftbox{0.3*#3};
   \coordinate (center) at (#1,#2);
   \coordinate (bottomright) at ($(center) + (+\shiftbox,-\shiftbox)$);
   \coordinate (topright)    at ($(center) + (+\shiftbox,+\shiftbox)$);
   \draw[linePlain] ($(bottomright) + (+\shift,-\shift)$) -- (bottomright);
   \draw[linePlain] (topright)    -- ($(topright)   + (+\shift,+\shift)$);
}

\newcommand{\arrowslefthalffull}[3]{
   \def\shift{0.3};
   \def\shiftbox{0.3*#3};
   \coordinate (center) at (#1,#2);
   \coordinate (bottomleft)  at ($(center) + (-\shiftbox,-\shiftbox)$);
   \coordinate (topleft)     at ($(center) + (-\shiftbox,+\shiftbox)$);
   \draw[lineWithArrowCenterEnd] (bottomleft)  -- ($(bottomleft) + (-\shift,-\shift)$);
   \draw[lineWithArrowCenterEnd] ($(topleft)     + (-\shift,+\shift)$) -- (topleft);
}

\newcommand{\arrowsrighthalffull}[3]{
   \def\shift{0.3};
   \def\shiftbox{0.3*#3};
   \coordinate (center) at (#1,#2);
   \coordinate (bottomright) at ($(center) + (+\shiftbox,-\shiftbox)$);
   \coordinate (topright)    at ($(center) + (+\shiftbox,+\shiftbox)$);
   \draw[lineWithArrowCenterEnd] ($(bottomright) + (+\shift,-\shift)$) -- (bottomright);
   \draw[lineWithArrowCenterEnd] (topright)    -- ($(topright)   + (+\shift,+\shift)$);
}

\newcommand{\threepointvertexleft}[4]{
   \draw[linePlain] (#2,#3+0.4*#4) -- (#2+0.6*#4,#3) -- (#2,#3-0.4*#4) -- (#2,#3+0.4*#4);
   \node at (#2+0.2*#4,#3) {#1};
}

\newcommand{\threepointvertexleftarrows}[4]{
   \threepointvertexleft{#1}{#2}{#3}{#4}
   \arrowslefthalffull{#2+0.4*#4}{#3}{4./3.*#4}
}

\newcommand{\threepointvertexright}[4]{
   \draw[linePlain] (#2,#3) -- (#2+0.6*#4,#3+0.4*#4) -- (#2+0.6*#4,#3-0.4*#4) -- (#2,#3);
   \node at (#2+0.4*#4,#3) {#1};
}

\newcommand{\threepointvertexrightarrows}[4]{
   \threepointvertexright{#1}{#2}{#3}{#4}
   \arrowsrighthalffull{#2+0.2*#4}{#3}{4./3.*#4}
}

\newcommand{\loopfullvertex}[4]{
   \def\shift{0.3*#4};
   \draw[#1] ($(#2,#3) + (\shift,\shift)$) .. controls ++(45:0.4) and ++(0:0.4) .. ($(#2,#3) + (0,0.6+0.3*#4)$) .. controls ++(180:0.4) and ++(135:0.4) .. ($(#2,#3) + (-\shift,\shift)$);
}

\newcommand{\fullvertex}[4]{
   \def\shift{0.3};   
   \def\shiftbox{0.3*#4};
   \coordinate (center) at (#2,#3);
   \coordinate (bottomleft)  at ($(center) + (-\shiftbox,-\shiftbox)$);
   \coordinate (topleft)     at ($(center) + (-\shiftbox,+\shiftbox)$);
   \coordinate (bottomright) at ($(center) + (+\shiftbox,-\shiftbox)$);
   \coordinate (topright)    at ($(center) + (+\shiftbox,+\shiftbox)$);
   \draw[linePlain, fill=verylightgray] (bottomleft) rectangle (topright);
   \node at (center) {#1};
}

\newcommand{\arrowslowerhalffullbis}[3]{
   \def\shift{0.3};
   \def\shiftbox{0.3*#3};
   \coordinate (center) at (#1,#2);
   \coordinate (bottomleft)  at ($(center) + (-\shiftbox,-\shiftbox)$);
   \coordinate (bottomright) at ($(center) + (+\shiftbox,-\shiftbox)$);
   \draw[linePlain] (bottomleft)  -- ($(bottomleft) + (-\shift,-\shift)$);
   \draw[linePlain] ($(bottomright) + (+\shift,-\shift)$) -- (bottomright);
}

\newcommand{\arrowslefthalf}[2]{
   \def\shift{0.3};
   \coordinate (center) at (#1,#2);
   \draw[lineWithArrowCenterEnd] (center)  -- ($(center) + (-\shift,-\shift)$);
   \draw[lineWithArrowCenterEnd] ($(center)     + (-\shift,+\shift)$) -- (center);
}

\newcommand{\arrowsrighthalf}[2]{
   \def\shift{0.3};
   \coordinate (center) at (#1,#2);
   \draw[lineWithArrowCenterEnd] ($(center) + (+\shift,-\shift)$) -- (center);
   \draw[lineWithArrowCenterEnd] (center)    -- ($(center)   + (+\shift,+\shift)$);
}

\newcommand{\barevertexwithlegs}[2]{
   \fill (#1,#2) circle (2pt) coordinate (center);
   \arrowslefthalf{#1}{#2}
   \arrowsrighthalf{#1}{#2}
}

\newcommand{\fullvertexwithlegs}[4]{
        \def\shift{0.3};  
        \def\shiftbox{0.3*#4};
        \coordinate (center) at (#2,#3);
        \coordinate (bottomleft)  at ($(center) + (-\shiftbox,-\shiftbox)$);
        \coordinate (topleft)     at ($(center) + (-\shiftbox,+\shiftbox)$);
        \coordinate (bottomright) at ($(center) + (+\shiftbox,-\shiftbox)$);
        \coordinate (topright)    at ($(center) + (+\shiftbox,+\shiftbox)$);
        \draw[linePlain] (bottomleft) rectangle (topright);
        \node at (center) {#1};
        \draw[lineWithArrowCenterEnd] (bottomleft) -- ($(bottomleft) + (-\shift,-\shift)$); 
         \draw[lineWithArrowCenterEnd] ($(topleft)     + (-\shift,+\shift)$) -- (topleft);
        \draw[lineWithArrowCenterEnd] ($(bottomright) + (+\shift,-\shift)$) -- (bottomright); 
        \draw[lineWithArrowCenterEnd] (topright) -- ($(topright) + (+\shift,+\shift)$);
}

\newcommand{\threepointvertexupper}[4]{
   \draw[linePlain] (#2-0.4*#4,#3+0.3*#4) -- (#2+0.4*#4,#3+0.3*#4) -- (#2,#3-0.3*#4) -- (#2-0.4*#4,#3+0.3*#4);
   \node at (#2,#3+0.1*#4) {#1};
}

\newcommand{\arrowsupperhalffull}[3]{
   \def\shift{0.3};
   \def\shiftbox{0.3*#3};
   \coordinate (center) at (#1,#2);
   \coordinate (topleft)     at ($(center) + (-\shiftbox,+\shiftbox)$);
   \coordinate (topright)    at ($(center) + (+\shiftbox,+\shiftbox)$);
   \draw[lineWithArrowCenterEnd] ($(topleft)     + (-\shift,+\shift)$) -- (topleft);
   \draw[lineWithArrowCenterEnd] (topright)    -- ($(topright)   + (+\shift,+\shift)$);
}

\newcommand{\threepointvertexupperarrows}[4]{
   \threepointvertexupper{#1}{#2}{#3}{#4}
   \arrowsupperhalffull{#2}{#3-0.1*#4}{4./3.*#4}
}

\newcommand{\threepointvertexlower}[4]{
   \draw[linePlain] (#2-0.4*#4,#3-0.3*#4) -- (#2+0.4*#4,#3-0.3*#4) -- (#2,#3+0.3*#4) -- (#2-0.4*#4,#3-0.3*#4);
   \node at (#2,#3-0.1*#4) {#1};
}

\newcommand{\arrowslowerhalffull}[3]{
   \def\shift{0.3};
   \def\shiftbox{0.3*#3};
   \coordinate (center) at (#1,#2);
   \coordinate (bottomleft)  at ($(center) + (-\shiftbox,-\shiftbox)$);
   \coordinate (bottomright) at ($(center) + (+\shiftbox,-\shiftbox)$);
   \draw[lineWithArrowCenterEnd] (bottomleft)  -- ($(bottomleft) + (-\shift,-\shift)$);
   \draw[lineWithArrowCenterEnd] ($(bottomright) + (+\shift,-\shift)$) -- (bottomright);
}

\newcommand{\threepointvertexlowerarrows}[4]{
   \threepointvertexlower{#1}{#2}{#3}{#4}
   \arrowslowerhalffull{#2}{#3+0.1*#4}{4./3.*#4}
}

\allowdisplaybreaks

\usepackage{subcaption}

\newcommand{\M}{{ \mathrm{M} }}
\newcommand{\D}{{ \mathrm{D} }}

\newcommand{\X}{{ \mathrm{X} }}

\definecolor{verylightgray}{RGB}{240,240,240}

\begin{document}

\begin{center}{\Large \textbf{
BosonFlow: A C++ codebase for dynamic fRG and single-boson exchange in correlated fermion systems
}}\end{center}

\begin{center}
Aiman Al-Eryani\textsuperscript{1}, Miriam Patricolo\textsuperscript{3},
Kilian Fraboulet\textsuperscript{2,3,4}
\end{center}

\begin{center}
{\bf 1} Theoretical Physics III, Ruhr-University Bochum, 44801 Bochum, Germany
\\
{\bf 2} Max-Planck-Institut für Festkörperforschung, Heisenbergstraße 1, 70569 Stuttgart, Germany
\\
{\bf 3} Institute of Information Systems Engineering, Vienna University of Technology, 1040 Vienna, Austria
\\
{\bf 4} Institut f\"ur Theoretische Physik and Center for Quantum Science, Universit\"at T\"ubingen, Auf der Morgenstelle 14, 72076 T\"ubingen, Germany
\\
* aiman.al-eryani@rub.de
\end{center}

\begin{center}
\today
\end{center}


\section*{Abstract}
{\bf
We present a unified C++ implementation of the functional renormalization group and the parquet equations within the single-boson exchange formalism for several paradigmatic tight-binding and impurity models at equilibrium. The implementation computes the full dynamic vertex and self-energies, with momentum dependence treated using a truncated unity framework. We implement multiple self-energy flow equations, cutoff schemes, and extensions ranging from the dynamical functional renormalization group to multiloop flow equations that incorporate cutoffs in both the propagator and the interaction. The codebase serves as a reference for recent developments in the fRG and parquet methods for correlated electron systems and provides a flexible foundation for developing new many-body approaches and extensions.
}

\vspace{10pt}
\noindent\rule{\textwidth}{1pt}
\tableofcontents\thispagestyle{fancy}
\noindent\rule{\textwidth}{1pt}
\vspace{10pt}

\section{Introduction}
\label{sec:Introduction}

The physics of strongly correlated electron systems is characterized by a rich variety of competing orders, where magnetic, charge-density-wave, and superconducting phases often emerge in close proximity. A proper theoretical understanding of these materials requires many-body methods that can treat different scattering instabilities in a diagrammatically unbiased way. Furthermore, the extended nature of low-energy excitations (i.e., the Fermi-surface) poses an additional formidable challenge, in that these low-energy excitations need to be captured by \emph{coupling functions} (with momentum and frequency dependencies) rather than coupling constants. Approaches based on the functional Renormalization Group (fRG)~\cite{Wetterich1993,Morris1994,Ellwanger1994,Berges2002,Delamotte2012,Metzner2012,Dupuis2021} and the parquet equations~\cite{ded64,ded64bis,Bickers1991,Bickers1992,BickersSelfConsistent2004} have emerged as powerful tools for this purpose, as they allow for the unbiased treatment of fluctuations across multiple channels of the vertex functions.

However, the numerical implementation of these methods is hindered by the high dimensionality of the two-particle vertex, which depends on three independent frequency and momentum arguments (for systems with translational invariance and energy conservation). In-so-far, researchers have been forced into a computational trade-off: neglecting frequency dependence to achieve high momentum resolution for lattice systems, or neglecting momentum dependence to resolve the full dynamics of impurity models. The neglect of frequency dependence carries profound qualitative implications; omitting vertex dynamics often leads to an unphysical overestimation of critical temperatures for phase transitions. More fundamentally, the dynamic (i.e., frequency-dependent) vertex and self-energy are essential for capturing quasi-particle lifetimes and non-Fermi-liquid behavior, such as the pseudogap phenomenon observed in the two-dimensional Hubbard model. Furthermore, resolving the frequency axis is required for a consistent treatment of retarded interactions, ensuring that the dynamic renormalization of mediators, such as phonons, is handled on an equal footing with electronic scattering.

The current ecosystem of open-source tools provides different strategies to address these challenges. High-performance libraries like divERGe~\cite{Profe2024} utilize the Truncated Unity (TU) framework~\cite{Husemann2009,Wang2012,Lichtenstein2017} to enable fRG calculations on arbitrary lattices and are excellent for the qualitative instability analysis of sophisticated many-orbital models. However, divERGe currently relies on a static vertex approximation, limiting its ability to resolve the vertex content on the frequency axis. Conversely, the KeldyshQFT~\cite{Ritz2024} codebase provides a sophisticated implementation of multiloop fRG and the self-consistent parquet equations directly in real frequencies. While powerful, KeldyshQFT is presently restricted to impurity models, such as the single-impurity Anderson model, as the inclusion of momentum dependence for lattice systems remains a significant computational hurdle. For lattice models that do include frequency dependencies, the Truncated Unity Parquet Solver (TUPS)~\cite{Eckhardt2020} provides a numerical iterative solution to the parquet equations using a standard parquet decomposition. Another recent work is the DiFfRG codebase~\cite{sattler2024diffrgdiscretisationframeworkfunctional}, which however is geared more towards high-energy applications that do not support a Fermi-surface.

In this work, we present \emph{\texttt{BosonFlow}}, a C++ codebase designed for quantitatively accurate calculations through the resolution of full dynamic vertex structures in lattice models. The framework serves both as a tool for controlled many-body calculations and as a reference and benchmark for future implementations, with a particular focus on the quantitative study of paradigmatic $SU(2)$-invariant single-orbital models. In contrast to TUPS, which employs a parquet decomposition, \texttt{BosonFlow} is based on the single-boson exchange (SBE) formulation~\cite{Krien2019} of the functional renormalization group~\cite{Bonetti2022,Fraboulet2022}. Combined with a truncated unity treatment of momentum dependence, this approach enables the computation of dynamic vertices and self-energies on the Matsubara axis.

The code provides a flexible environment for exploring different approximations, self-energy flow equations and cutoff schemes. It includes, as an option, the multiloop fRG framework~\cite{Kugler2018a,Kugler2018b,Kugler2018c,Gievers2022,Fraboulet2025}, which removes the unphysical regulator dependence characteristic of conventional one-loop truncations. In addition, \texttt{BosonFlow} offers tools for detailed vertex and self-energy diagnostics and includes an interpolating flow for DMF$^2$RG~\cite{Taranto2014,Vilardi2019}, enabling the incorporation of correlated starting points for the fRG flow based on dynamical mean-field theory~\cite{Metzner1989,Georges1996,Rohringer2018}.

The framework further incorporates several recent methodological developments, including the $G$+$U$ flow scheme~\cite{AlEryani2025b}, in which both bare propagators and bare interactions are dressed with regulators. This extends the multiloop fRG approach and notably allows for efficient temperature  scans for systems involving electron-phonon couplings.

\texttt{BosonFlow} supports calculations at finite temperature for a broad class of models, both in the thermodynamic limit and on finite lattices. It involves notably the two-dimensional square-lattice Hubbard model~\cite{Hubbard1963,Kanamori1963,Gutzwiller1963,Zhang1989}, a paradigmatic model used, e.g., for the description of high-temperature superconductivity in cuprates~\cite{Anderson1987,Zhang1988,Qin2022,Arovas2022}. \texttt{BosonFlow} also includes the Hubbard atom~\cite{Hubbard1963} and chain~\cite{Lieb1968}, the two-dimensional triangular-lattice Hubbard models~\cite{Hubbard1963,Kanamori1963,Gutzwiller1963,DosSantos1993,Sahebsara2008}, and Anderson-type impurity systems~\cite{Anderson1961}. Extensions involving phonons are also available, such as the two-dimensional Hubbard--Holstein model~\cite{Holstein1959,Freericks1995} and the Anderson--Holstein impurity model~\cite{ELagos2001,Hewson2002}. In addition, the code features a self-consistent SBE-parquet solver, providing an alternative to flow-based approaches and a platform for benchmarking dynamic vertex effects in correlated electron systems.

\section{Fermionic functional renormalization group and its single-boson exchange implementation}

To study competing orders in fermionic systems, the most common fRG implementation is the one-loop fRG~\cite{Morris1994,Metzner2012}, whose flow equations are depicted diagrammatically in Fig.~\ref{fig:1loopfRGequations}. This scheme is in practice limited to weak or intermediate couplings. However, extensions have been designed to treat strong-coupling regimes. In that respect, one can mention the DMF$^2$RG~\cite{Taranto2014,Vilardi2019,Bonetti2022,Katanin2019}, which merges DMFT with fRG in such a way that the fRG equations are solved using DMFT results as initial conditions. Furthermore, to improve the accuracy of the fermionic fRG beyond the conventional one-loop implementation, one can also consider the multiloop fRG~\cite{Kugler2018a,Kugler2018b,Kugler2018c} that introduces loop corrections derived from the Bethe-Salpeter equations~\cite{Salpeter1951}.

\begin{figure*}[t!]
    \centering
        \begin{align*}
            \scalebox{1.2}{${\displaystyle\partial_{\Lambda}\textcolor{blue}{\Sigma} }$} \hspace{0.15cm} &\scalebox{1.1}{${\displaystyle = }$} \hspace{0.1cm} \scalebox{1.1}{\tikzm{fRGhierarchy1forSBE}{
		\fullvertex{\scalebox{1.1}{${\displaystyle \hspace{0.015cm} \textcolor{blue}{V}}$}}{0}{0}{1}
		\arrowslowerhalffullbis{0}{0}{1}
		\loopfullvertex{linePlain}{0}{0}{1}
            \draw[linePlain] (0.0+0.1,0.96+0.15-0.07) -- (0.0-0.1,0.96-0.15-0.07);
		\node[above] at (0.0,0.98-0.07) {\scalebox{0.75}{${\displaystyle\textcolor{black}{\partial_\Lambda G^{-1}_{0,\Lambda} }}$}};
	}}\\
            \scalebox{1.2}{${\displaystyle\partial_{\Lambda} \textcolor{blue}{V} }$} \hspace{0.15cm} &\scalebox{1.1}{${\displaystyle = }$} \hspace{0.1cm} \scalebox{1.1}{\tikzm{fRGhierarchy2forSBE}{
				\arrowslefthalffullbis{0}{0}{1}
				\fullvertex{\scalebox{1.1}{${\displaystyle \hspace{0.035cm} \textcolor{blue}{V}}$}}{0}{0}{1}
				\draw[linePlain] (0.3,0.3) to [out=45, in=135] (1.02,0.3);
				\draw[linePlain] (0.759,0.445) -- (0.761,0.445);
				\draw[linePlain] (1.02,-0.3) to [out=225, in=315] (0.3,-0.3);
				\draw[linePlain] (0.561,-0.445) -- (0.559,-0.445);
				\draw[linePlain] (0.66+0.1,0.44+0.15) to (0.66-0.1,0.44-0.15);
				\node[above] at (0.66,0.46) {\scalebox{0.75}{${\displaystyle\textcolor{black}{\partial_\Lambda G^{-1}_{0,\Lambda} }}$}};
				\fullvertex{\scalebox{1.1}{${\displaystyle \hspace{0.035cm} \textcolor{blue}{V}}$}}{1.32}{0}{1}
				\arrowsrighthalffullbis{1.32}{0}{1}
			}}
        \end{align*}
    \caption{Schematic representation of the flow equations underlying the one-loop fRG. These equations represent the evolution of vertex functions (the self-energy $\Sigma$ and the two-particle vertex $V$) with a flow parameter $\Lambda$ introduced in the bare propagator of the theory $G_0$ ($G_0\rightarrow G_{0,\Lambda}$). The squares represent the two-particle vertex $V$, whereas the solid black lines correspond to the full propagator $G$ and the dashes denote insertions of the derivatives $\partial_\Lambda G_{0,\Lambda}^{-1}$, as indicated on the diagrams.}
    \label{fig:1loopfRGequations}
\end{figure*}

The generic form of the flow equations underlying the aforementioned fermionic fRG approaches (one-loop or multiloop fRG, DMF$^2$RG or weak-coupling fRG) for lattice models (which rely on translational invariance and energy conservation) is: 
\begin{subequations}
\begin{align}
    \partial_\Lambda \Sigma(k_1) &= \mathbf{RHS}_\Sigma[k_1;\Sigma,V; \dot{\Sigma}, \dot{V}], \label{eq:FlowEqSigma}\\
    \partial_\Lambda V(k_1,k_2,k_3) &= \mathbf{RHS}_V[k_1,k_2,k_3;\Sigma,V; \dot{\Sigma}, \dot{V}], \label{eq:FlowEqV}
\end{align}
\label{eq:GenericFlowEqs}
\end{subequations}
where the external indices include spatial momenta and Matsubara frequencies, i.e. $k_i=(\mathbf{k}_i,\nu_i)$ ($i=1,2,3$). Here, $\Sigma$ and $V$ are respectively the self-energy and the two-particle vertex of the studied system. The latter is essential for the study of competing instabilities as susceptibilities can be extracted from $V$ via, e.g., Bethe-Salpeter equations. The functionals $\mathbf{RHS}_\Sigma$ and $\mathbf{RHS}_V$ involve integrals/sums over momenta and frequencies, with integrands depending on both $\Sigma$ and $V$. Although not specified explicitly in Eqs.~\eqref{eq:GenericFlowEqs}, we note that these integrands also depend on $\partial_\Lambda G_{0,\Lambda}^{-1}$ (within the one-loop fRG) or $\partial_\Lambda G^{-1} = \partial_\Lambda G_{0,\Lambda}^{-1}-\partial_\Lambda\Sigma$ (within the multiloop fRG), where $G_{0,\Lambda}$ is the regulated bare propagator which defines the flow scheme (since the flow parameter $\Lambda$ is originally introduced via the substitution $G_0\rightarrow G_{0,\Lambda}$ in conventional fRG approaches\footnote{Within the $G$+$U$ flow scheme mentioned previously~\cite{AlEryani2025b}, a regulator is also introduced in the bare interaction $U$ (via the substitution $U\rightarrow U_\Lambda$) such that the multiloop flow equations involve both $\partial_\Lambda G^{-1}$ and $\partial_\Lambda U$ in that case.}).

The differential equations~\eqref{eq:GenericFlowEqs} are therefore integro-differential equations, and their accurate treatment requires specific numerical approaches in the \emph{C++ \texttt{BosonFlow} codebase} presented in this article. As mentioned previously, \texttt{BosonFlow} can treat one-loop and multiloop fRG equations, within DMF$^2$RG but also within conventional flow schemes. These ``conventional flow schemes'' refer to the $\Omega$-flow~\cite{Husemann2009}, the temperature flow~\cite{Honerkamp2001} and the interaction flow~\cite{Honerkamp2004}, where the flow parameter $\Lambda$ corresponds to a frequency, a temperature scale and the interaction strength, respectively. As opposed to DMF$^2$RG, these conventional flow schemes rely on uncorrelated starting points for the flow with initial conditions inferred, e.g., from the non-interacting or the classical theory.

The heart of the numerical treatment of Eqs.~\eqref{eq:GenericFlowEqs} is the description of the two-particle vertex $V(k_1,k_2,k_3)$, which is a highly complicated object. For the lattice models considered in \texttt{BosonFlow}, it depends on three spatial momenta and three frequencies, i.e., $V=V(k_1,k_2,k_3)$ where the momenta $k_i=(\mathbf{k}_i,\nu_i)$ ($i=1,2,3$) contain each a spatial momentum and a frequency. In practice, the vertex $V$ is thus often decomposed into several objects such that the differential equation~\eqref{eq:FlowEqV} is in practice replaced by other flow equations. In that respect, the C++ \texttt{BosonFlow} codebase uses the SBE decomposition of $V$~\cite{Krien2019}, which is shown in Fig.~\ref{fig:SBEdecomposition}.

\begin{figure*}[t!]
\begin{align*}
\tikzm{SBEdecomposition_V}{
			\fullvertexwithlegs{$V$}{0}{0}{1}
			\node[above, align=center] at (-0.7,0.58) {\footnotesize $2$};
			\node[above, align=center] at (0.7,0.58) {\footnotesize $2^\prime$};
			\node[below, align=center] at (-0.7,-0.53) {\footnotesize $1^\prime$};
			\node[below, align=center] at (0.7,-0.58) {\footnotesize $1$};
		}
		& = \ \tikzm{SBEdecomposition_nablaph}{
				\threepointvertexupperarrows{$\overline{\lambda}_{ph}$}{0}{0.82}{1.4}
				\bosonfull{0}{0.4}{0}{-0.4}
				\node at (0.4,0) {$w_{ph}$};
				\threepointvertexlowerarrows{$\lambda_{ph}$}{0}{-0.82}{1.4}
				\node[left, align=center] at (-0.7,1.8) {\footnotesize $2$};
				\node[right, align=center] at (0.7,1.8) {\footnotesize $2^\prime$};
				\node[left, align=center] at (-0.7,-1.75) {\footnotesize $1^\prime$};
				\node[right, align=center] at (0.75,-1.775) {\footnotesize $1$};
			}
		\ + \ \tikzm{SBEdecomposition_nablaphx}{\threepointvertexleftarrows{$\overline{\lambda}_{\overline{ph}}$}{-0.18}{0}{1.4}
				\bosonfull{0.66}{0}{1.46}{0}
				\node at (1.06,0.3) {$w_{\overline{ph}}$};
				\threepointvertexrightarrows{$\lambda_{\overline{ph}}$}{1.46}{0}{1.4}
				\node[above, align=center] at (-0.61,0.85) {\footnotesize $2$};
				\node[above, align=center] at (2.73,0.85) {\footnotesize $2^\prime$};
				\node[below, align=center] at (-0.61,-0.84) {
                \footnotesize $1^\prime$};
				\node[below, align=center] at (2.73,-0.85) {
                 \footnotesize $1$};
			} \ + \ \tikzm{SBEdecomposition_nablapp}{
				\threepointvertexleft{$\overline{\lambda}_{pp}$}{-0.18}{0}{1.4}
				\bosonfull{0.66}{0}{1.46}{0}
				\node at (1.06,0.3) {$w_{pp}$};
				\node at (1.06,0.74) {};
				\threepointvertexright{$\lambda_{pp}$}{1.46}{0}{1.4}
				\draw[lineWithArrowCenterEnd] (-0.18,-0.56) -- (-0.48,-0.86);
				\draw[lineWithArrowCenterEnd] (2.6,-0.86) -- (2.3,-0.56);
				\draw[lineWithArrowCenterEnd] (-0.18,0.56) to [out=20, in=180] (2.6,0.86);
				\draw[lineWithArrowCenterStart] (-0.48,0.86) to [out=0, in=160] (2.3,0.56);
				\node[above, align=center] at (-0.61,0.85) {\footnotesize $2$};
				\node[above, align=center] at (2.73,0.85) {\footnotesize $2^\prime$};
				\node[below, align=center] at (-0.61,-0.84) {
                \footnotesize $1^\prime$};
				\node[below, align=center] at (2.73,-0.85) {\footnotesize $1$};
			} \\
            & \phantom{=} 
		\ + \sum_{r=ph,\overline{ph},pp}\tikzm{SBEdecomposition_M}{
			\fullvertexwithlegs{$M_r$}{0}{0}{1}
			\node[above, align=center] at (-0.7,0.58) {\footnotesize $2$};
			\node[above, align=center] at (0.7,0.58) {\footnotesize $2^\prime$};
			\node[below, align=center] at (-0.7,-0.53) {\footnotesize $1^\prime$};
			\node[below, align=center] at (0.7,-0.58) {\footnotesize $1$};
		} \ + \ \tikzm{SBEdecomposition_I2PI}{
			\fullvertexwithlegs{$\mathcal{I}^{\text{2PI}}$}{0}{0}{1.35}
			\node[above, align=center] at (-0.8,0.68) {\footnotesize $2$};
			\node[above, align=center] at (0.8,0.68) {\footnotesize $2^\prime$};
			\node[below, align=center] at (-0.8,-0.63) {\footnotesize $1^\prime$};
			\node[below, align=center] at (0.8,-0.68) {\footnotesize $1$};
		}
            \ -3 \ 
            \tikzm{SBE_decomposition_Gamma0}{
			\barevertexwithlegs{0}{0}
                \node[above, align=center] at (-0.4,0.28) {\footnotesize $2$};
			\node[above, align=center] at (0.4,0.28) {\footnotesize $2^\prime$};
			\node[below, align=center] at (-0.4,-0.23) {\footnotesize $1^\prime$};
			\node[below, align=center] at (0.4,-0.28){\footnotesize $1$};
		}
        \end{align*}
\caption{Single-boson exchange decomposition of the two-particle vertex $V$. Each diagram correspond to a four-point object with respect to spatial momenta $\mathbf{k}_{\alpha}$, Matsubara frequencies $\nu_{\alpha}$ and spin indices $\sigma_{\alpha}$, i.e., $\alpha=(\mathbf{k}_{\alpha},\nu_{\alpha},\sigma_{\alpha})$ with $\alpha=1,2,1^\prime,2^\prime$. The last term $-3 U$ is a double-counting correction that removes the bare interaction $U$ included by convention as leading order to each single-boson exchange contribution of the first line.}
\label{fig:SBEdecomposition}
\end{figure*}

\begin{table}[t!]
\centering
\renewcommand{\arraystretch}{1.6}

\begin{tabular}{|c|c|c|c|}
\hline
\diagbox[width=1.6cm,height=1.2cm]{\scriptsize Param.}{$r$\ \ } & \textbf{$ph$} & \textbf{$\overline{ph}$} & \textbf{$pp$} \\
\hline

$\mathbf{Q}_r$ 
& $\mathbf{k}_{1} - \mathbf{k}_{1'}$
& $\mathbf{k}_1 - \mathbf{k}_{2'}$
& $\mathbf{k}_1 + \mathbf{k}_2$ \\

$\Omega_r$
& $\nu_1 - \nu_{1'}$
& $\nu_1 - \nu_{1'}$
& $\nu_1 + \nu_2$ \\
\hline

$\mathbf{k}_r$ 
& $\mathbf{k}_2$
& $\mathbf{k}_2$
& $\mathbf{k}_2$ \\

$\nu_r$
& $\nu_2 + \lfloor \Omega_{ph}/2 \rfloor$
& $\nu_2 + \lfloor \Omega_{\overline{ph}}/2\rfloor$
& $\nu_2 - \lceil \Omega_{pp}/2 \rceil$ \\
\hline

$\mathbf{k}_r'$ 
& $\mathbf{k}_{1'}$
& $\mathbf{k}_{2'}$
& $\mathbf{k}_{2'}$ \\

$\nu_r'$
& $\nu_{1'}+ \lfloor \Omega_{ph}/2 \rfloor$
& $\nu_{2'} + \lfloor \Omega_{\overline{ph}}/2 \rfloor$
& $\nu_{2'} - \lceil \Omega_{pp}/2 \rceil$ \\
\hline
\end{tabular}

\caption{Conventions used in \texttt{BosonFlow} for the spatial momenta and the Matsubara frequencies for the two-particle vertex $V$, and for the objects $w_r$, $\lambda_r$ and $M_r$ introduced within its SBE decomposition (see Fig.~\ref{fig:SBEdecomposition}), in each diagrammatic channel $r$ (with $r=ph,\overline{ph},pp$). Note that $\lceil ... \rceil$ and $\lfloor ... \rfloor$ round their argument up and down to the nearest bosonic Matsubara frequency, respectively.}
\label{tab:FrequencyMomentumParametrization}
\end{table}

Within the SBE framework, part of the contributions to the vertex $V$ is parametrized in terms of bosonic propagators $w_r$ and fermion-boson vertices $\lambda_r$ (also called Hedin vertices) and their conjugates $\overline{\lambda}_r$, where $r$ corresponds to either the \emph{particle-hole} ($ph$), \emph{particle-hole-crossed} ($\overline{ph}$) or the \emph{particle-particle} ($pp$) channels. These three channels originally label the two-particle-reducible (2PR) diagrams of $V$ depending on the nature of the two propagator lines\footnote{By definition, the diagrams assigned to the $ph$, $\overline{ph}$ and $pp$ channels contain two transverse antiparallel, two antiparallel and two parallel lines that can be cut to split them into two disconnected parts. See Refs.~\cite{KuglerPhDthesis,Gievers2022,AlEryani2025b,Fraboulet2025} for more details on this point.} that can be cut to split the diagram into two disconnected parts. For the SBE decomposition, this translates into single-boson exchange contributions with bosons with different momenta, i.e., $k_1-k_{1^\prime}$, $k_1-k_{2^\prime}$ and $k_1+k_{2}$ in the $ph$, $\overline{ph}$ and $pp$ channels, respectively. It is thus natural to introduce different \emph{momentum and frequency conventions} for objects related to one of these three channels. Those used in \texttt{BosonFlow} are defined by Tab.~\ref{tab:FrequencyMomentumParametrization}.

Only part of the 2PR diagrams contributing to $V$ can be interpreted as single-boson exchange processes, which correspond to the so-called \emph{$U$-reducible diagrams} that can be split into two disconnected parts by cutting a bare interaction $U$. The remaining 2PR contributions correspond to multiboson exchange processes and are assigned to the multiboson rest functions $M_r$. Finally, the diagrams that are two-particle-irreducible (i.e., not 2PR) are encompassed in the irreducible vertex $I^{\mathrm{2PI}}$. We note that, within both the one-loop and multiloop fRG approaches, the \emph{parquet approximation} is used for conventional flow schemes\footnote{Note the exception of the DMF$^2$RG where it is set to its DMFT value, i.e., $I^{\mathrm{2PI}}=I^{\mathrm{2PI}}_{\mathrm{DMFT}}$.}, which implies that $I^{\mathrm{2PI}}$ is set to the bare interaction of the studied model ($I^{\mathrm{2PI}}=U$). Hence, within the SBE approach, the one-loop and multiloop fRG rely on flow equations for bosonic propagators, fermion-boson vertices and multiboson rest functions. The heart of the motivation for this substitution is that bosonic propagators $w_r$ and fermion-boson vertices $\lambda_r$ all possess a simpler frequency and momentum structure compared to the original vertex $V(k_1,k_2,k_3)$. We indeed have $w_r(Q_r)$ and $\lambda_r(Q_r,k_r)$ for $r=ph,\overline{ph},pp$, with $Q_r=(\mathbf{Q}_r,\Omega_r)$ and $k^{(\prime)}_r=(\mathbf{k}^{(\prime)}_r,\nu^{(\prime)}_r)$ (see Tab.~\ref{tab:FrequencyMomentumParametrization}). The multiboson rest functions $M_r(Q_r,k_r,k_r^{(\prime)})$ however still depend on three momenta and three frequencies, so the efficiency of the SBE approach depends on our ability to neglect these objects. The quality of this approximation has been studied within the one-loop and the multiloop fRG in Refs.~\cite{Fraboulet2022,Fraboulet2025}.

It should also be noted that, for models involving retarded or non-local\footnote{``Non-local'' interactions refer here to interactions between electrons on different sites.} interactions, the bare interaction exhibits non-trivial momentum or frequency dependencies, which induces more involved momentum or frequency structures of $w_r$ and $\lambda_r$. To retain the efficiency of the SBE approach, a redefinition of bosonic propagators, fermion-boson vertices and multiboson rest functions was proposed in Refs.~\cite{AlEryani2024,AlEryani2025} based on a diagrammatic criterion called $\mathcal{B}$-reducibility. This amounts to splitting the bare interaction $U$ into a bosonic part $\mathcal{B}_r$ and a fermionic part $\mathcal{F}_r$:
\begin{equation}
    U(Q_r,k_r,k^\prime_r) = \mathcal{B}_r(Q_r) + \mathcal{F}_r(Q_r,k_r,k^\prime_r).
    \label{eq:BplusFsplitting}
\end{equation}
The bosonic bare interaction $\mathcal{B}_r$ thus contain all terms depending only on the bosonic momentum of their respective channel. The bosonic propagators $w_r$ and fermion-boson vertices $\lambda_r$ can then be used to parametrize only the diagrams that can be split into two disconnected parts by cutting a bosonic interaction $\mathcal{B}_r$, hence the notion of $\mathcal{B}$-reducibility. The bosonic propagators, fermion-boson vertices and multiboson rest functions considered in \texttt{BosonFlow} are based on those generalized definitions. Furthermore, each of these vertices are four-point objects with respect to spin indices, $w_{r;\sigma_{1^\prime}\sigma_{2^\prime}|\sigma_1\sigma_2}$, $\lambda_{r;\sigma_{1^\prime}\sigma_{2^\prime}|\sigma_1\sigma_2}$ and $M_{r;\sigma_{1^\prime}\sigma_{2^\prime}|\sigma_1\sigma_2}$. Since \texttt{BosonFlow} focuses on $SU(2)$-spin-symmetric systems only a few of those components are actually finite and independent. By selecting the relevant spin components in the $ph$, $\overline{ph}$ and $pp$ channels~\cite{Gievers2022,AlEryani2025b,Fraboulet2025}, one can introduce the so-called physical channels, coined as magnetic ($\mathrm{M}$), density ($\mathrm{D}$) and superconducting ($\mathrm{SC}$). Each of these new channels have their own bosonic propagator, fermion-boson vertex and multiboson rest function in such a way that the SBE scheme allows for replacing Eqs.~\eqref{eq:GenericFlowEqs} by the following differential equations:
\begin{subequations}
\begin{align}
    \partial_\Lambda \Sigma(k) &= \mathbf{RHS}_{\Sigma}[k;\Sigma,w_{\mathrm{X}},\lambda_{\mathrm{X}},M_{\mathrm{X}}], \label{eq:FlowEqSigmaSBE}\\
    \partial_\Lambda w_{\mathrm{X}}(Q) &= \mathbf{RHS}_{w_{\mathrm{X}}}[Q;\Sigma,w_{\mathrm{X}},\lambda_{\mathrm{X}},M_{\mathrm{X}}], \label{eq:FlowEqw}\\
    \partial_\Lambda \lambda_{\mathrm{X}}(Q,k) &= \mathbf{RHS}_{\lambda_{\mathrm{X}}}[Q,k;\Sigma,w_{\mathrm{X}},\lambda_{\mathrm{X}},M_{\mathrm{X}}], \label{eq:FlowEqlambda}\\
    \partial_\Lambda M_{\mathrm{X}}(Q,k,k^\prime) &= \mathbf{RHS}_{M_{\mathrm{X}}}[Q,k,k^\prime;\Sigma,w_{\mathrm{X}},\lambda_{\mathrm{X}},M_{\mathrm{X}}], \label{eq:FlowEqM}
\end{align}
\label{eq:GenericFlowEqsSBE}
\end{subequations}
for $\mathrm{X}=\mathrm{M},\mathrm{D},\mathrm{SC}$. Eqs.~\eqref{eq:GenericFlowEqsSBE} correspond to the generic form of the flow equations treated in \texttt{BosonFlow}. We stress again that the objects $w_{\mathrm{X}}$, $\lambda_{\mathrm{X}}$ and $M_{\mathrm{X}}$ are obtained from specific spin components of $w_r$, $\lambda_r$ and $M_r$ and their exact definitions can be found in Refs.~\cite{AlEryani2025b,Fraboulet2025}. The exact form of Eqs.~\eqref{eq:GenericFlowEqsSBE} can be found in Ref.~\cite{Fraboulet2022} for the one-loop fRG. For the multiloop fRG, dependenc on derivatives is also included in the right hand-side. Their expressions can be found in Ref.~\cite{Fraboulet2025}, and in Ref.~\cite{AlEryani2025b} for the extension based on the $G$+$U$ flow scheme.

\section{Applications}

\texttt{BosonFlow} has already been applied to several correlated-electron models. It started with an analysis of the two-dimensional square-lattice Hubbard model with the one-loop fRG~\cite{Fraboulet2022}, focused on the merits of the SBE formulation. This study focused on the weak-coupling regime regime of this model, at temperatures above the pseudo-critical antiferromagnetic transition of the one-loop fRG. This transition is an artifact of the one-loop approximation, and produces a long-range order in a two-dimensional system at finite temperature that is in principle forbidden by the Mermin--Wagner theorem~\cite{mer66,hoh67,col73bis}. An important finding reported in Ref.~\cite{Fraboulet2022} is that, in the vicinity of the pseudo-critical transition, only the magnetic bosonic propagator $w_{\mathrm{M}}(Q)$ exhibits diverging behavior whereas the other objects appear to remain finite. This makes it easier to describe the sharp momentum features inherent to diverging behavior, considering that bosonic propagators $w_{\mathrm{X}}(Q)$ have a much simpler momentum dependence compared to the original two-particle vertex $V(k_1,k_2,k_3)$. Furthermore, Ref.~\cite{Fraboulet2022} also finds that the contribution of the multiboson rest function appears to be negligible in most of the studied regimes (weak couplings for temperatures above the pseudo-critical transition), which implies that $V(k_1,k_2,k_3)$ is efficiently described by the simpler objects $w_{\mathrm{X}}(Q)$ and $\lambda_{\mathrm{X}}(Q,k)$. These conclusions highlights the advantage of the SBE formulation of the (one-loop) fRG and lay the foundation for subsequent work based on \texttt{BosonFlow}.

The natural next step consists in performing similar investigations within the multiloop fRG, which is know to provide quantitatively for two-dimensional lattice systems at weak coupling~\cite{Hille2020b}. Furthermore, multiloop fRG results coincide with the parquet approximation at loop convergence~\cite{TagliaviniHille2019,Kugler2018a}, which is known to be consistent with the Mermin--Wagner theorem~\cite{Bickers1992,AlEryani2026}. In that sense, the multiloop fRG cures the artificial long-range order produced by the one-loop approximation. The multiloop fRG equations in the SBE framework were originally developed in diagrammatic channels in Ref.~\cite{Gievers2022}. Subsequent projects based on \texttt{BosonFlow} were carried out by reformulating the multiloop SBE fRG equations in physical channels~\cite{Patricolo2025,Fraboulet2025}, which provide a compact formulation for $SU(2)$-spin-symmetric systems. First, self-energy flow equations were derived from the Schwinger-Dyson equation for each physical channel ($\mathrm{X}=\mathrm{M},\mathrm{D},\mathrm{SC}$) and rewritten explicitly in terms of bosonic propagators and fermion-vertices~\cite{Patricolo2025}, thus providing a simpler reformulation of the Schwinger-Dyson equation in previous multiloop fRG studies~\cite{Hille2020a,Hille2020b}. Ref.~\cite{Patricolo2025} also analyzed in particular how the different channel representations of the Schwinger-Dyson equation thus obtained are affected by form-factor truncations. This was achieved by studying the pseudogap opening in the two-dimensional square-lattice Hubbard model, with numerical results produced with \texttt{BosonFlow}. As a next step, the multiloop fRG equations for bosonic propagators, fermion-boson vertices and multiboson rest functions were reformulated also in terms of physical channels in Ref.~\cite{Fraboulet2025}. These equations were implemented in \texttt{BosonFlow} and an analysis of the corresponding numerical results showed that the parquet approximation is accurately reproduced at loop convergence without multiboson rest functions. This highlights that, within the multiloop fRG, the two-particle vertex $V(k_1,k_2,k_3)$ can also be efficiently described by simpler objects within the SBE framework.

The aforementioned studies were performed for the conventional two-dimensional square-lattice Hubbard model. For application to models with extended or non-local interactions that respect the role of the rest function as a residual of essential bosonic fluctuations, the SBE had to be generalized in terms of the $\mathcal{B}+\mathcal{F}$-splitting of the bare interaction mentioned previously with Eq.~\eqref{eq:BplusFsplitting}. This generalization is built in \texttt{BosonFlow}, and was applied to thoroughly study the extended Hubbard and Hubbard--Holstein models within the one-loop fRG in Refs.~\cite{AlEryani2024,AlEryani2025}, with SBE-inspired fluctuation diagnostics. More recently, the work of Ref.~\cite{AlEryani2025b} generalized the multiloop fRG within the SBE framework with the $G$+$U$ flow scheme introduced in Sec.~\ref{sec:Introduction}, thus allowing for the implementation of temperature flows for systems with electron-phonon couplings. In that context, \texttt{BosonFlow} has been employed to produce benchmark calculations for the Anderson impurity model and a temperature-flow application to the Anderson--Holstein impurity model, an impurity model involving an electron-phonon coupling. 

In a recent clarification of the question of whether multiloop fRG and the parquet approximation satisfy the Hohenberg-Mermin-Wagner theorem, self-consistent calculations with \texttt{BosonFlow} were also used to perform numerical fits for the anomalous dimension~\cite{AlEryani2026}.

\section{Usage}

Performing a calculation with \texttt{BosonFlow} is fairly straightforward. After cloning and setting up the github repository in \url{https://github.com/AG-Andergassen/BosonFlow.git} -- see repository page for more details -- and ensuring the required dependencies are installed, the user:
\begin{enumerate}
    \item Sets \emph{technical parameters} and flags (solution strategy, approximations, details on frequency boxes and momentum meshes, approximations ..etc)   within the file \directory{config.mk}.
    \item Builds the code by executing e.g. \texttt{make run -j}. A resulting binary is then produced. 
    \item Executes the built binary with desired giving an output directory and further \emph{model parameters} as command line arguments.
\end{enumerate}
Resulting observables and flowing objects are outputted in \texttt{HDF5} files. 

The following subsections summarize the required software dependencies, some of the important compile-time options and the available runtime parameters, and the structure of the generated output. More details can be found in \texttt{BosonFlow}'s github webpage.

\subsection{Installation and Prerequisites}

\texttt{BosonFlow} requires a compiler that supports C++20 (gcc~$\geq 13$ or LLVM~$\geq 16$) together
with the external libraries listed in Table~\ref{tab:dependencies}. On HPC clusters these
are typically loaded via an environment manager (such as \texttt{modules}) or package managers (such as \texttt{spack}).

\begin{table}[h]
\centering
\caption{Required external dependencies.}
\label{tab:dependencies}
\begin{tabular}{lll}
\hline
\textbf{Library} & \textbf{Minimum version} & \textbf{Purpose} \\
\hline
Boost   & 1.66  & ODE solver (odeint) and utilities \\
FFTW3   & 3.3.8 & Fast Fourier transforms for bubble convolutions \\
HDF5 & 1.10.3 & Persistent storage and results of calculations \\
Eigen3  & 3.3.7 & Linear algebra operations and matrices \\
OpenMP  & ---   & Multi-threaded parallelization (recommended) \\
\hline
\end{tabular}
\end{table}

The standard build procedure, which produces an optimised multi-threaded binary at
\texttt{./bin/run}, is
\begin{lstlisting}[language=sh]
make clean && make run -j
\end{lstlisting}
which builds with \texttt{gcc}. Additional build targets are available: \texttt{make runclang} compiles with Clang/LLVM and
often improves code generation, while \texttt{make dbg} and \texttt{make dbgclang} produce debug builds with
bounds checking and threading disabled, suitable for use with the debuggers \texttt{gdb} and \texttt{lldb} respectively. We have found that Clang/LLVM produces the more optimized binaries. 

\subsection{Technical parameters and Model Selection}
All compile-time flags are specified in \texttt{config.mk}. These specify various technical parameters of the calculations. For clarity, it is useful to group them into: (i) flow and solver choices, (ii) physical model selection, (iii) SBE-specific flags and approximation, (iv) multiloop corrections, (v) grid parameters and performance flags. Within some groups the flags are mutually exclusive; only one must be enabled at a time, while in others they act as complementary tools.

\subsubsection{Core flow and solver choices}
The most consequential decisions concern the solver strategy and the choice of the flow scheme, as these shape the entire numerical approach. The \textbf{solver strategy} is mutually exclusive: \texttt{FLOW\_EQUATION\_METHOD} performs an ODE-based fRG flow, while \texttt{SELF\_CONSISTENT\_METHOD} uses solves the SBE equations by fixed point iteration~\cite{Krien2021}. The choice of a \textbf{$G$-flow scheme} is likewise mutually exclusive and determines the type of regulator/cutoff implemented on the propagator; the available options are \texttt{OMEGA\_FLOW} introduced in Ref.~\cite{Husemann2009}, \texttt{INT\_FLOW} introduced in Ref.~\cite{Honerkamp2004}, \texttt{TEMP\_FLOW} introduced in Ref.~\cite{Honerkamp2001}. More details can also be found in Ref.~\cite{Fraboulet2025}. Additionally and independently, one may specify a $U$-flow scheme which dictate how the the bare interaction will flow, as introduced in Ref.~\cite{AlEryani2025b}. The \texttt{ACTUAL\_U
\_FLOW} scheme implements a simple interpolation of the bare interaction from $0$ to its full value, whereas the \texttt{FREE\_U\_FLOW} allows for more sophisticated model-dependent deformation of the interaction. In that case, one needs to implement additional functions in the model class. Currently, the \texttt{FREE\_U\_FLOW} flag is usable if used together with the \texttt{TEMP\_FLOW} flag with the Anderson impurity Holstein model, which implements a correct temperature flow even though the model has a retarded interaction. See Ref.~\cite{AlEryani2025b} for more details. One may also turn on only the \texttt{ACTUAL\_U\_FLOW} flag without any of the $G$-flow flags.

The self-energy treatment can be chosen as \texttt{SELFEN\_FLOW} for the standard one-loop flow in the flow equation solution strategy, or as \texttt{SELFEN\_SDE\_SBE\_$^*$} for a channel-specific SDE/SBE~\cite{PatricoloInprep}, where $^*$ stands for one of the magnetic, density, or superconducting channels for any of the two solution strategies. If none of these flags are chosen, then the self-energy is neglected. When doing a self-consistent calculation, the \texttt{SELFEN\_FLOW} flag is unused, and one must employ one of the three SDE forms of the self-energy.   
Table~\ref{tab:compileflags-solver} summarises the mutually exclusive strategy flags, and Table~\ref{tab:compileflags-tuning} covers other tuning and resolution options.

\begin{table}[h]
\centering
\caption{Strategy and formulation flags (mutually exclusive within each group).}
\label{tab:compileflags-solver}
\begin{tabular}{llp{7cm}}
\hline
\textbf{Group} & \textbf{Flag} & \textbf{Effect} \\
\hline
Solver
  & \texttt{FLOW\_EQUATION\_METHOD}    & ODE-based fRG integration \\
  & \texttt{SELF\_CONSISTENT\_METHOD}  & Fixed-point iteration \\
\hline
$G$-flows
  & \texttt{OMEGA\_FLOW}                    & Frequency regulator \\
  & \texttt{INT\_FLOW}                      & Interaction flow (through $G$) \\
  & \texttt{TEMP\_FLOW}                     & Temperature flow \\
$U$-flows
  & \texttt{ACTUAL\_U\_FLOW}                    & Interaction flow (through $U$)  \\
  & \texttt{FREE\_U\_FLOW}                      & Interaction flow (freely defined through model)\\
\hline
Self-energy
  & \texttt{SELFEN\_FLOW}                        & Standard one-loop flow \\
  & \texttt{SELFEN\_SDE\_SBE\_MAGNETIC}          & SDE/SBE, magnetic channel \\
  & \texttt{SELFEN\_SDE\_SBE\_DENSITY}           & SDE/SBE, density channel \\
  & \texttt{SELFEN\_SDE\_SBE\_SUPERCONDUCTING}   & SDE/SBE, SC channel \\
\hline
\end{tabular}
\end{table}

\begin{table}[h!]
\centering
\caption{Some relevant resolution and run-control flags. }
\label{tab:compileflags-tuning}
\begin{tabular}{lp{8.5cm}}
\hline
\textbf{Flag} & \textbf{Effect} \\
\hline
\texttt{K\_DIM\_VAL}                       & Number of momentum grid points in each dimension \\
\texttt{COUNT\_VAL}                        & Frequency grid box multiplier \\
\texttt{FORMFACTOR\_SHELL\_COUNT\_VAL}     & Shell count for form-factor models; set to 1 for atom/impurity models \\
\texttt{P\_IN\_K\_VAL}                  & Multiplier for a finer momentum grid in each dimension, for the integration of the bubble functions \\
\texttt{STATIC\_CALCULATION}           & Neglects frequencies of the vertex and the self-energy \\
\texttt{MULTILOOP} & Utilize multiloop RHS\\
\texttt{EXTRA\_LOOP\_NUM\_VAL}             & When multiloop RHS is used, specifies additional vertex loops (0 = 1-loop+Katanin, 1 = 2-loop, \ldots) \\
\texttt{SBEa\_APPROXIMATION}              & Enables the SBEa approximation by neglecting the rest-function flow in flow-equation mode \\
\texttt{SBEb\_APPROXIMATION}              & Rest function is neglected prior to computing the flow equations \\
\texttt{SET\_MIXED\_BUBBLES\_TO\_ZERO}     & Approximates bubble components off-diagonal in form-factors to zero; reducing computational cost \\
\texttt{FIX\_FILLING}                       & Pin filling to its initial value by self-consistently adjusting the chemical potential \\
\texttt{PRECOMPUTE\_STATE\_PROJECTIONS}    & Precomputes form-factor projections; trades memory for speed \\
\texttt{UTILIZE\_ALGEBRAIC\_SYMMETRIES}   & Exploits algebraic (spin/frequency) symmetries to reduce the number of independently evaluated indices; highly recommended \\
\texttt{UTILIZE\_LATTICE\_SYMMETRIES}     & Exploits lattice point-group symmetries to further reduce the independent index set; highly recommended \\
\texttt{MULTIFILE\_OUTPUT}                 & Writes output to multiple files (one per flow step) \\
\texttt{AUTORESUME\_CALCULATION}           & Restarts calculation from the last outputted step \\
\texttt{MICKY\_MOUSE}                      & Debug mode with minimal grid parameters; not for production \\
\hline
\end{tabular}
\end{table}
\subsubsection{Choosing a model}
The physical model is selected through the macro
\texttt{THE\_MODEL\_VAL}, which accepts the identifiers listed in
Table~\ref{tab:models}. Lattice models include on-site interaction $U$, nearest-neighbour
hopping $t$, and optional longer-range terms. Holstein variants add electron--phonon
coupling characterized by a coupling strength $g_0$ and a phonon frequency $\omega_0$.
The long-range variant
adds a screened Coulomb interaction, while the Anderson impurity models couple a (singly) impurity site to a
bath defined by a hybridization function. For an overview of model parameters, pass the \texttt{--help} option as a command line argument to the built binary.

\begin{table}[h!]
\centering
\caption{Available models for \texttt{THE\_MODEL\_VAL}. Models that have been tested thoroughly are marked with ``\checkmark''.}
\label{tab:models}
{
\setlength{\tabcolsep}{5pt}
\begin{tabular}{ll}
\hline
\textbf{Identifier} & \textbf{Description} \\
\hline
\texttt{HubbardAtom} \checkmark              & Single-site Hubbard atom \\
\texttt{AndersonImpurity}  \checkmark        & Anderson impurity with hybridisation bath \\
\texttt{AndersonImpurityHolstein}\checkmark  & Anderson impurity + Holstein phonons \\
\texttt{SquareHubbard}  \checkmark           & Square-lattice Hubbard model \\
\texttt{SquareHubbardHolstein} \checkmark    & Square Hubbard + Holstein phonons \\
\texttt{SquareHubbardLongRange}& Square Hubbard + screened Coulomb \\
\texttt{TriangularHubbard}    & Triangular-lattice Hubbard model \\
\texttt{ChainHubbard}         & 1D chain Hubbard model \\
\texttt{FCCHubbard}           & FCC-lattice Hubbard model \\
\hline
\end{tabular}
}
\end{table}

\subsubsection{SBE-specific flags and approximations}

Several compile-time flags are specifically relevant within the treatment of the flow equations in the SBE formalism. In particular, enabling \texttt{SBEa\_APPROXIMATION} disables the \text{flow} of the multiboson rest function. The flag \texttt{SBEb\_APPROXIMATION} goes further by eliminating multi-boson contributions entirely. In the self-consistent solution strategy however, both flags identically leave out the computation of the rest function. Notably, whereas both approximations maintain the same computational reduction, the SBEa approximation still contains partial differentiated multi-boson contributions implicit through the flow of the fermion-boson vertices, and so remains a superior approximation available only through fRG; see Ref.~\cite{Fraboulet2022}.

Several additional flags allow for further approximations, such as \texttt{NO\_HEDIN\_VERTEX\_FLOW}, which freezes the fermion-boson vertex at its bare value and disables its flow, or 

\noindent 
\texttt{SET\_MIXED\_BUBBLES\_TO\_ZERO} which forces mixed bubble cross terms to zero. The 

\noindent
\texttt{PRECOMPUTE\_STATE\_PROJECTIONS} flag when enabled precomputes the form-factor projection matrices trading memory for performance. 

\subsubsection{Multiloop calculations}

Solving the multiloop fRG flow equations is done by enabling the \texttt{MULTILOOP} flag. Internally, this switches to a multiloop RHS class which can handle the calculation of the multiloop corrections. Then, \texttt{EXTRA\_LOOP\_NUM\_VAL} specifies the number of extra loop corrections beyond the base scheme.


When the standard \texttt{SELFEN\_FLOW} flag is selected, the self-energy is computed from the conventional one-loop flow equation using the four-point vertex. Alternatively, one of the Schwinger--Dyson variants, \texttt{SELFEN\_SDE\_SBE\_MAGNETIC}, \texttt{SELFEN\_SDE\_SBE\_DENSITY}, or \texttt{SELFEN\_SDE\_SBE\_SUPERCONDUCTING}, may be chosen. In that case, the self-energy is reconstructed from the corresponding Schwinger--Dyson equation within the single-boson-exchange formalism.

The parameter \texttt{EXTRA\_LOOP\_NUM\_VAL} determines the highest loop correction to be included. For \texttt{EXTRA\_LOOP\_NUM\_VAL}=0, only the one-loop contribution together with the Katanin correction is retained. For \texttt{EXTRA\_LOOP\_NUM\_VAL}=1, the two-loop vertex correction is added. Values \texttt{EXTRA\_LOOP\_NUM\_VAL}$\ge 2$ include three-loop and higher-loop corrections systematically.

When \texttt{EXTRA\_LOOP\_NUM\_VAL}$>0$, the code performs a self-consistent inner loop at each RG step in order to iteratively refine the self-energy together with the multiloop vertex corrections. Details can be found in Ref.~\cite{Fraboulet2025}.

Put briefly, each RG step proceeds as follows. First, the one-loop right-hand sides for the vertex and self-energy are evaluated. The propagators are then updated using the Katanin correction, and the relevant particle-particle and particle-hole bubble diagrams are computed and cached. The algorithm subsequently enters a self-consistency loop, in which the multiloop vertex corrections are constructed from these bubbles and, if applicable, the self-energy is updated according to the selected Schwinger--Dyson equation. After each update, convergence is checked against \texttt{ABS\_ERROR\_SELFENERGY\_ITERATIONS\_VAL} defined in \texttt{config.mk}. The iteration is repeated until convergence is reached or until the maximum number of iterations specified by \texttt{SELFENERGY\_ITERATIONS\_MAX\_VAL} is attained. The flow then advances to the next RG scale.

At each loop order $\ell \ge 2$, the code monitors convergence using the thresholds 

\noindent
\texttt{ERR\_LOOP\_ABS\_VAL} and \texttt{ERR\_LOOP\_REL\_VAL}. By default, the multiloop calculation terminates early once the norm of the current loop correction falls below the prescribed threshold. This behavior can be disabled with \texttt{FORCE\_CALCULATE\_ALL\_MULTILOOP\_CORRECTIONS}, which forces the evaluation of all loop orders up to \texttt{EXTRA\_LOOP\_NUM\_VAL}. This option is mainly useful for diagnostics and convergence studies.

Finally, we note that the Katanin correction mentioned above is enabled by default with the \texttt{MULTILOOP} flag at all loop orders, but it can be omitted by enabling the \texttt{NO\_KATANIN} flag found in \texttt{config.mk}.

\subsubsection{Momentum and frequency grids}

\begin{table}
    \centering
    \footnotesize
    \begin{tabular}{c|cccc}
Object & \parbox{45pt}{Bosonic\\frequency}& \parbox{42pt}{Fermionic\\frequency}& \parbox{50pt}{Bosonic\\momentum}& \parbox{58pt}{Fermionic momentum\\/ form factors}\\
 \\
 \hline
 \\
         $\Sigma$, $\dot{\Sigma}$ &  $\varnothing$&  $20(\texttt{COUNT\_VAL})$& $\varnothing$ & $(\texttt{K\_DIM\_VAL})^d$\\
         $\Pi$, $\dot{\Pi}$ &  $128(\texttt{COUNT\_VAL})+1$&   $128(\texttt{COUNT\_VAL})$& $(\texttt{K\_DIM\_VAL}\times \texttt{PX\_IN\_KX\_VAL})^d$ & ff.\\
         $w$, $\dot{w}^{(\ell)}$ &  $128 (\texttt{COUNT\_VAL}) + 1$&  $\varnothing$& $(\texttt{K\_DIM\_VAL})^d+$ refinement & $\varnothing$\\
         $\lambda$, $\dot{\lambda}^{(\ell)}$ &  $4(\texttt{COUNT\_VAL}) + 1$&  $4(\texttt{COUNT\_VAL})$& $(\texttt{K\_DIM\_VAL})^{d}+$ refinement & ff.\\
         $M$, $\dot{M}^{(\ell)}$ &  $4(\texttt{COUNT\_VAL}) + 1$&  $2(\texttt{COUNT\_VAL})$& $(\texttt{K\_DIM\_VAL})^d+$ refinement & ff. \\
    \end{tabular}
    \caption{Frequency, momentum, and form-factor resolution used in the parametrization of the objects in relation to the technical parameters in \texttt{config.mk}. Within truncated-unity~\cite{Lichtenstein2017,Husemann2009}, bosonic transfer momenta are discretized on Brillouin-zone grids, while fermionic momentum dependence is represented through a small form-factor basis (``ff.''). For the self-energy, fermionic momenta are resolved directly on the coarse $(\texttt{K\_DIM\_VAL})^d$ grid.}
    \label{tab:technical_parameters}
\end{table}

The numerical resolution is determined primarily by \texttt{K\_DIM\_VAL}, \texttt{COUNT\_VAL}, \texttt{PX\_IN\_KX\_VAL}, and \texttt{FORMFACTOR\_SHELL\_COUNT\_VAL}. These control the momentum discretisation, Matsubara-frequency resolution, and form-factor shell size, respectively.

More specifically, \texttt{K\_DIM\_VAL} sets the number of points per dimension of the coarse momentum grid used for vertex and self-energy parametrization. In addition, the code uses a refined momentum grid with \texttt{K\_DIM\_VAL}$\times$\texttt{PX\_IN\_KX\_VAL} points per dimension. This finer grid enters bubble integrations (see the $\Pi,\dot{\Pi}$ row in Table~\ref{tab:technical_parameters}), where particle-particle and particle-hole bubbles develop increasingly sharp structures near the Fermi surface as the temperature is lowered.  
Additional refinement at selected local points in the Brillouin zone can be specified through command line arguments to the binary.

Table~\ref{tab:FrequencyMomentumParametrization} summarizes, for each object, which frequency boxes are used and whether momenta are represented on the coarse bosonic grid, the refined bubble grid, or through fermionic form factors (``ff.''). This reflects the truncated-unity representation, where fermionic momentum dependence is projected onto a compact form-factor basis. The entries marked ``$+$ refinement'' indicate the supplemented by the aforementioned additional refinement points.
 
 The parameter \texttt{FORMFACTOR\_SHELL\_COUNT\_VAL} further specifies the number of form-factor shells included in the truncated unity treatment of secondary fermionic momenta. For models on the square lattice, magic values such as $1.3$ indicate including only three: $s$-wave + two $p$-wave form factors, whereas $1.5$ includes only two: the $s$-wave and the $d$-wave form factors.  For models without spatial structure, such as the Hubbard atom or Anderson impurity, \texttt{FORMFACTOR\_SHELL\_COUNT\_VAL} is taken to be $1$ (i.e. only the on-site form factor is included). 

\subsubsection{Example representative calculation}
A representative \texttt{config.mk} configuration for a square-Hubbard $1\ell$+Katanin fRG calculation with a frequency cutoff function involves the following

\begin{lstlisting}[language=make]
#...
CFLAGS += -D FLOW_EQUATION_METHOD 
CFLAGS += -D THE_MODEL_VAL=SquareHubbard
CFLAGS += -D COUNT_VAL=5
CFLAGS += -D K_DIM_VAL=16
CFLAGS += -D P_IN_K_VAL=5
CFLAGS += -D FORMFACTOR_SHELL_COUNT_VAL=1.0
CFLAGS += -D OMEGA_FLOW
CFLAGS += -D SELFEN_FLOW
CFLAGS += -D MULTIFILE_OUTPUT 
CFLAGS += -D EXTRA_LOOP_NUM_VAL=0
CFLAGS += -D SPLIT_SUSC_CONTRIBUTIONS
#...

\end{lstlisting}
The specified resolution is sufficient to perform calculations at inverse temperature $\beta t= 10$. This will lead additionally to a run performing fluctuation diagnostics of the susceptibility (as done e.g. in Ref.~\cite{AlEryani2025}) and to output this in the observables. 

\subsection{Runtime Parameters and Execution}

The compiled binary is invoked as

\begin{lstlisting}[language=sh]
./bin/run output_directory [options]
\end{lstlisting}

where \directory{output_directory} is a positional argument specifying a directory where the resulting HDF5 files will be
written. All physical parameters carry sensible
model-dependent defaults, Table~\ref{tab:runtimeparams}
lists the most commonly used runtime options.
\begin{table}[h]
\centering
\caption{Selected runtime parameters and their applicability.}
\label{tab:runtimeparams}
{
\setlength{\tabcolsep}{5pt}
\begin{tabular}{lp{6.6cm}p{3.4cm}}
\hline
\textbf{Option} & \textbf{Description} & \textbf{Models} \\
\hline
\texttt{--beta FLOAT}             & Inverse temperature $\beta = 1/T$         & All \\
\texttt{--uint FLOAT}             & On-site interaction $U$                   & All \\
\texttt{--mu FLOAT}               & Chemical potential $\mu$                  & All \\
\texttt{--filling FLOAT}          & Target filling used for initial chemical-potential adjustment; if omitted, filling follows $\mu$ & All \\
\texttt{--u-prime FLOAT}             & Nearest-neighbour interaction $U'$         & Square,Triangular, Chain \\
\texttt{--t-prime FLOAT}          &  hopping $t'$                          & Square,Triangular, FCC \\
\texttt{--t-prime-prime FLOAT}    & hopping $t''$                        & Triangular, FCC \\
\texttt{--g0 FLOAT}               & Electron--phonon coupling $g_0$           & Holstein variants \\
\texttt{--omega0 FLOAT}           & Phonon frequency $\omega_0$               & Holstein variants \\
\texttt{--D FLOAT}                & Bath half-bandwidth                       & Anderson impurity \\
\texttt{--delta0 FLOAT}           & Hybridisation strength $\Delta_0$         & Anderson impurity variants \\
\texttt{--dos-type <BOX|CONST>} & type of Hybridisation function & Anderson impurity variants\\
\texttt{--refine-at label:r}      & Refine grid near symmetry point           & Lattice models \\
\texttt{--help}                   & Print options and defaults                & All \\
\hline
\end{tabular}
}
\end{table}
Example executions for the square Hubbard model are

\begin{lstlisting}[language=sh]
./bin/run ./results/run1 --beta 10.0 --uint 3.0 --mu -1.0
./bin/run ./results/run2 --beta 8.0 --uint 2.5 \
          --refine-at idx_00:0.02 --refine-at idx_pipi:0.015
\end{lstlisting}

The \texttt{--refine-at} option accepts a \texttt{label:radius} format, where
\texttt{label} identifies a Brillouin-zone symmetry point (e.g.\ \texttt{idx\_00} for
$\Gamma = (0,0)$, \texttt{idx\_pipi} for $M = (\pi,\pi)$, \texttt{idx\_ppi\_0} for
$X = (\pi,0)$) and \texttt{radius} $\in [0, 1]$  specifies the refinement region (in percent) in momentum space.
This allows targeted improvement of momentum resolution near points of physical interest,
such as van Hove singularities or ordering wavevectors.

For parameter scans, independent runs can be launched in parallel with a simple shell
loop:

\begin{lstlisting}[language=sh]
for U in 1.0 2.0 3.0 4.0 5.0; do
    mkdir -p "results/U_${U}"
    ./bin/run "./results/U_${U}" --beta 10.0 --u "$U" --mu -0.5 &
done
wait
\end{lstlisting}

\subsection{Output Structure and Data Access}
\texttt{BosonFlow} writes its results in HDF5 format. With \texttt{MULTIFILE\_OUTPUT} enabled, the output directory contains a sequence of numbered snapshot files (\texttt{0.h5}, \texttt{1.h5}, \ldots), which store intermediate states of the flow at different scales, and \texttt{Params.h5}, which collects the main parameters used for the run. If the vertex grows unboundedly during the flow, signaling either a physical instability or divergence in the fixed point iteration, the solver stops and writes a file with the suffix \texttt{\_DIVERGENT.h5}. When \texttt{SPLIT\_SUSC\_CONTRIBUTIONS} is enabled, the susceptibility output is further decomposed into individual channel contributions, which facilitates identification of the dominant fluctuation mode. Compile-time flags also affect the output filenames by encoding parameter choices used in the run.

The HDF5 file is organized into the groups \texttt{Sig}, \texttt{lambda\_func}, \texttt{w\_func}, and \texttt{Flow\_obs}. The group \texttt{Sig} stores the self-energy on the fermionic frequency and momentum grids, together with the corresponding grids in \texttt{fgrid} and \texttt{momgrid}. The real and imaginary parts are stored separately as \texttt{Sig/RE} and \texttt{Sig/IM}.

The group \texttt{lambda\_func} contains the channel-resolved fermion-boson vertices in the density, magnetic, and superconducting channels. Their real and imaginary parts are stored separately as \texttt{RE\_D}, \texttt{IM\_D}, \texttt{RE\_M}, \texttt{IM\_M}, \texttt{RE\_SC}, and \texttt{IM\_SC}. The associated bosonic and fermionic frequency grids and the momentum grid are stored in \texttt{bgrid}, \texttt{fgrid}, and \texttt{momgrid}.

The group \texttt{w\_func} stores the corresponding bosonic propagators, again resolved into density, magnetic, and superconducting channels and split into real and imaginary parts. Its associated bosonic frequency and momentum grids are stored in \texttt{w\_func/bgrid} and \texttt{w\_func/momgrid}.

Postprocessed two-particle observables are collected in \texttt{Flow\_obs}, which includes polarization functions in \texttt{Postprocessing\_Polarisation\_info}, susceptibilities in 

\noindent\texttt{Postprocessing\_Susc\_info}, $s$-wave projected susceptibilities in \texttt{S\_Wave\_Susc\_info}, and decompositions of the fermion-boson vertices in \texttt{Postprocessing\_Lambda\_info}. For all these quantities, real and imaginary parts are stored separately using the prefixes \texttt{RE\_} and \texttt{IM\_}. For example, the magnetic susceptibility is stored as \texttt{RE\_Susc\_m} and \texttt{IM\_Susc\_m}.

The momentum grids are also stored explicitly as datasets named \texttt{momgrid}. These grids are represented as two-component momentum points, while the frequency grids are stored separately as \texttt{fgrid} for fermionic frequencies and \texttt{bgrid} for bosonic frequencies.

HDF5 files can be read in most popular languages (e.g. Python, MATLAB, or Julia). A minimal Python to read the real and imaginary parts of the self-energy using the \texttt{h5py} module is
\begin{lstlisting}[language=Python,basicstyle=\scriptsize\ttfamily]
import h5py
import matplotlib.pyplot as plt

with h5py.File('results/final.h5', 'r') as f:
    fermionic_frequency_grid = f['Sig/fgrid']
    sigma_real = f['Sig/RE'][:,0,0,0] #frequency_idx, momentum_idx, spin_idx, spin_idx
    sigma_imaginary = f['Sig/IM'][:,0,0,0] # 0 momentum idx always corresponds to zero momentum
    plt.plot(fermionic_frequency_grid, sigma_real, "x-")
    plt.plot(fermionic_frequency_grid, sigma_imaginary, "+-")
    plt.show()
\end{lstlisting}
\subsubsection{Momentum- and Frequency-Resolved Plots}
\label{subsection:plots}
Beyond direct inspection of individual datasets, the HDF5 output can be easily used to extract one-dimensional cuts from the full multidimensional data. In momentum-resolved plots, one usually does not display an observable against the raw momentum-point index, but instead evaluates it along a physically meaningful path through the Brillouin zone, typically a high-symmetry path such as $\Gamma$-X-M-$\Gamma$. The corresponding sequence of momentum-point indices can be read from \texttt{Params.h5}, matched to the momentum grid stored in \texttt{final.h5}, and converted into a one-dimensional coordinate given by the cumulative distance along the path. The observable is then evaluated at each point of this trajectory, for example at fixed bosonic Matsubara frequency $\Omega=0$ in the case of a static susceptibility. This representation makes the momentum structure of the quantity directly visible and allows one to identify the wave vectors at which fluctuations are strongest.

Frequency-resolved plots can be constructed analogously by fixing a momentum point and evaluating the observable on the Matsubara-frequency grid. For the self-energy, for example, one typically selects representative momentum points, such as the antinodal point $(\pi,0)$ and the nodal point $(\pi/2,\pi/2)$, and then plots either $\mathrm{Re}\,\Sigma(\nu,\mathbf{k})$ or $\mathrm{Im}\,\Sigma(\nu,\mathbf{k})$ as a function of the Matsubara frequency $\nu$.
However, since the momentum-resolved self-energy is stored on a discrete Brillouin-zone grid, the mapping between array index and physical momentum is read from the dataset \texttt{Sig/momgrid}, whose entry \texttt{momgrid[i]} gives the pair $(k_x,k_y)$ associated with momentum index $i$. Nodal and antinodal cuts are then obtained by selecting the indices whose momentum coordinates correspond to the desired points of the Brillouin zone.
An example of a momentum-resolved postprocessing workflow and the corresponding plot is shown below.
\begin{lstlisting}[language=Python,basicstyle=\scriptsize\ttfamily]
import h5py
import numpy as np
import math
import matplotlib.pyplot as plt

# Read final output and parameter file
with h5py.File("results/final.h5", "r") as f, h5py.File("results/Params.h5", "r") as p:

    # Momentum grid from final.h5
    momgrid = np.array(f["w_func/momgrid"])

    # High-symmetry path Gamma-X-M-Gamma from Params.h5
    path_indices = np.array(p["Model/Special_paths/path_Gamma_X_M"])

    # Construct the cumulative distance along the path
    path_momenta = momgrid[cleaned_path]
    xaxis = np.zeros(len(path_momenta))
    for i in range(1, len(path_momenta)):
        dk = path_momenta[i] - path_momenta[i - 1]
        xaxis[i] = xaxis[i - 1] + math.sqrt(dk[0]**2 + dk[1]**2)

    # Example: static magnetic susceptibility along the path
    chi_m_re = np.array(f["Flow_obs/Postprocessing_Susc_info/RE_Susc_m"])
    omega_index = 0
    chi_path = chi_m_re[omega_index, cleaned_path]

# Plot
plt.plot(xaxis, chi_path, marker="o", linewidth=0.8)
plt.xticks(
    [0.0, math.pi, 2.0 * math.pi, 2.0 * math.pi + math.sqrt(2.0 * math.pi**2)],
    [r"$\Gamma$", "X", "M", r"$\Gamma$"]
)
plt.ylabel(r"$\chi^{\mathrm M}(\mathbf{Q},\Omega=0)$")
plt.xlabel(r"momentum path")
plt.tight_layout()
plt.show()
\end{lstlisting}

\clearpage
\section{Implementation details}

In this section, we give a technical overview of the implementation details and architecture of the code. A lot can be learned by inspecting \directory{src/start.cpp} (where the \cpp{main} function of the C++ program is located) and \directory{include/start.h}.   The code starts by initializing the model, grids, symmetry containers, flow-scheme, input-output (I/O), state and a right-hand side (RHS) object. The model layer specifies the lattice geometry and bare interactions; the frequency-scheme layer constructs Matsubara grids, form factors, and projection matrices; symmetry initialization and interpolation setup jointly reduce the number of required evaluations. The flow-scheme layer then registers the regulator choice and propagator functions. Observable tracking is configured, and finally the solver (either ODE integration or self-consistent iteration) is executed, with the persistent HDF5 output recording the evolving self-energy $\Sigma$ and SBE vertex components $w_\X, \lambda_X, M_\X$ alongside derived observables.

\subsection{Overview and High-Level Design}
A central design principle we followed in \texttt{BosonFlow} is the \emph{Separation of Concerns} (SoC) principle, where the program is divided into distinct parts, each responsible for certain aspect of the program. In particular, the three main physics inputs --- the
model, the regulator, and the solution method --- have implementations which details are abstracted from each other and from the state and RHS implementations. 

\begin{enumerate}
\item \textbf{The model} is primarily responsible for the lattice geometry and interactions.
Model classes define the lattice, the Brillouin zone, the
single-particle dispersion $\epsilon_\mathbf{k}$, and the bare two-particle
interaction in both the full-vertex and SBE-channel forms. They carry no
knowledge of the flow scheme, the frequency grid layout, or how the
equations are solved. Their static \cpp{Init()} method populates
all geometry data structures once at startup; after that, every other
part of the code calls model class members through static class members.
Model-specific parameters (such as a Hubbard $U$, next-nearest neighbor hopping $t'$, electron-phonon coupling $g_0$, phonon frequency $\omega_0$,
\ldots etc) are set through
compile-time macro values that are set
at startup from the command-line arguments. 

Nearly every class in the codebase is a class template parametrized by a
\cpp{Model} type. This type argument propagates
model-specific compile-time constants --- spatial dimension
(\cpp{Model::dim}), number of momentum points, the number of form-factors, and array sizes derived from
\verb|K_DIM_VAL|, \verb|COUNT_VAL|, and
\verb|FORMFACTOR_SHELL_COUNT_VAL| respectively --- into state containers,
RHS functors, bubble arrays, symmetry maps, truncated unity interchannel projection matrices, I/O
routines and so on.

\item \textbf{The flow scheme} (through the \cpp{fRGFlowScheme<Model>} class) is
responsible only for the choice and implementation of the regulators. It implements the scale-dependent bare propagator $G_{0,\Lambda}$ for each of the available schemes. In addition, it implements further related objects such as the dressed propagator $G_\Lambda$ and the single-scale propagator $S_\Lambda$. A suitably re-scaled ``physical'' propagator is also specified that can be used, for example, to calculate the filling at intermediate stages of the flow. For some cutoff schemes, tail-correction asymptotics used in bubble
frequency sums are also implemented. The choice of scheme --- such as those listed in table~\ref{tab:compileflags-solver} --- is registered once at startup via the \cpp{UseFlow(G_FlowSchemeName::*)} method of the class in the \cpp{main} function of the program. Optionally, a flowing bare
interaction $U_\Lambda$ can be enabled independently through specifying a
\cpp{U_FlowSchemeName::*} argument, resulting in the addition of $\dot{\mathcal{B}}$ and $\dot{\mathcal{F}}$ contributions to the
flow equations without changing anything else. In principle, there is no hard limitation to what combination of $G$- and $U$-flow schemes can be used together. The only requirement is that they share the same parametrization of the flow parameter $t$.

\item \textbf{The solver} choice is determined largely by the choice of the RHS (\cpp{rhs_*_t<Model, state_t>} class ) and the state (\cpp{state_*_t<Model>} class) classes. If one enables the flag
\cpp{FLOW_EQUATION_METHOD}, an ODE solver calls \cpp{rhs_*_t::operator(state, ..., t)} at each step of the flow and passes the
returned derivative state to the adaptive step-size controller. On the other hand, for
\cpp{SELF_CONSISTENT_METHOD}, where the SBE equations are solved self-consistently, in each iteration \cpp{vertex(...)} and
\cpp{selfenergy(...)} methods are called to update the vertex and self-energy components separately to allow finer control over vertex and self-energy convergence. Crucially, the same
\cpp{state_*_t}, \cpp{observer_*_t}, and I/O classes are used in both cases --- the solver choice
affects only the iteration logic in \directory{src/start.cpp}.
\end{enumerate}
This three-way separation means that extending any of the three --- whether adding a new
lattice model, a new cutoff scheme, or a new solution strategy/RHS --- does not require touching the other components, enabling ``modularity'' of the code.

\subsection{Models}
All major components of the code are templated on a \cpp{Model}
type chosen at compile time by the \cpp{THE_MODEL_VAL}
preprocessor flag set in \cpp{config.mk}, see Table~\ref{tab:models}. 

A new model is created by writing a class that inherits from \cpp{Hubbard<dim>},
where \cpp{dim} is the spatial dimension. The \cpp{Hubbard<dim>} itself inherits from an \cpp{AbstractModel} class, from which more general models can be written. The model is selected at compile time
via the \cpp{THE_MODEL_VAL} preprocessor flag; once set, the entire rest of the
code is templated on that type and model-specific data --- spatial dimension,
coarse and fine momentum counts, form-factor count --- propagates automatically
into the state containers, RHS functors, symmetry maps, and I/O routines, without
any further changes.

\subsubsection{Initialization}
A concrete model's \cpp{Init()} method takes no arguments and is responsible for
building all geometry. It constructs the lattice basis vectors (\cpp{basis_t<dim>}),
the point-group matrices grouped into conjugacy classes, and optionally a map of
named special momentum points and paths used for output. These objects, together with an
optional \cpp{refine_at_points} map for grid refinement and optional extra form
factors, are forwarded to \cpp{Hubbard<dim>::Init(...)}:
\begin{lstlisting}[language=C++]
Hubbard<dim>::Init(basis, point_group, special_points_coords,
                   special_paths_coords, refine_at_points, extra_form_factors);
\end{lstlisting}
This function handles each of the following constructions: the real-space lattice, the coarse
and fine Brillouin-zone meshes, the point-group action in momentum-index space,
the form-factor container, form-factor symmetry maps, Fourier-transform weights,
and the coarse-to-fine and fine-to-coarse momentum-index maps. Thus, implementing a new lattice model does not require re-specifying any of these objects: one only needs to inherit from
\cpp{Hubbard<dim>} and call its \cpp{Init()} function from the model's own
\cpp{Init()} method. 
Other metadata such as the human-readable name and a list of
parameter--value pairs for the output file are provided via the static methods
\cpp{GetName()} and \cpp{GetParamNameValuePairs()}.

\subsubsection{Indices, not coordinates}
An important design principle that we followed is that \emph{all} model-level functions are
called with discrete integer indices; the rest of the code never handles actual
momentum coordinates. Outisde of the model class, a momentum argument is a flat index \cpp{idx_K} into the coarse momentum-space (e.g. Brillouin zone)
mesh; a fermionic frequency argument is an index \cpp{idx_w} into the Matsubara
grid; a form-factor argument is a flat index \cpp{idx_m}. When an actual
coordinate is genuinely needed inside the model, it is recovered locally
via \cpp{GetMomentumCoordFromCoarseIdx(idx_K)} for coarse-grid momenta or
\cpp{GetMomentumCoordFromFineIdx(idx_p)} for fine-grid momenta. Coordinate arithmetic such as adding two momenta or taking the negative is likewise performed similar member functions of the model class, which which take indices as arguments and return indices corresponding to the momentum point that result from the operation.  This abstraction
is what allows the same flow equations to run on a zero-dimensional impurity and
a two-dimensional lattice model without any modification at the level of the RHS. This highlights one way in which the SoR principle manifests itself in the code design.

\subsubsection{Model couplings}
Each model is additionally characterized by its bare couplings; namely the dispersion, which defines its bare propagator $G_0$, and the bare vertex, expressed through the splitting $\mathcal{B}_\X + \mathcal{F}_\X$.
\paragraph{Dispersion.}
The single-particle dispersion is specified through two functions. The primary
one, \cpp{E(int idx_p, int idx_w)}, takes a fine-grid momentum index and a
Matsubara frequency index and returns the bare dispersion. Although most models are
frequency-independent and ignore \cpp{idx_w}, the argument is included in the
signature so that models coupled to a bath (such as the Anderson-impurity model) can be
supported without changing any calling code. In practice, models precompute the
dispersion on the fine momentum grid during \cpp{Init()} and let \cpp{E(...)}
perform a simple lookup\footnote{A second function, \cpp{E_of_p_coord(coord_t<dim>)},
takes an actual momentum coordinate rather than an index; it is used only in debugging branches of the code, where bubbles are computed using adaptive continuous integration over the
Brillouin zone rather than a sum over grid points.}.

\paragraph{Bare two-particle interaction.}
The model provides the bare interaction in several representations corresponding
to the different objects in the SBE channel decomposition.

In the model implementation, it is specified by defining
\begin{lstlisting}[language=C++]
static double vertex_4pt_bare(
    int idx_w1_in, int idx_w2_in, int idx_w1_out, int idx_w2_out,
    int idx_p1_in, int idx_p2_in, int idx_p1_out, int idx_p2_out,
    int s1_in, int s2_in, int s1_out, int s2_out);
\end{lstlisting}
which takes all four fermionic frequency and fine-grid momentum indices together
with spin/orbital labels\footnote{Although, as of yet, the code does not support multiband models or models that are not $SU(2)$-spin invariant, so the last four arguments are currently ignored in all currently available models. The code also does not use the fourth incoming/outgoing arguments since these would otherwise be fixed by momentum and energy conservation. }. Even here, no coordinates appear; if the physical formula
involves a transferred momentum $\mathbf{K}$, the model computes the
corresponding index internally using model methods such as \cpp{SumFineMomentaIdxes(...)} and
\cpp{GetNegativeFineMomentumIdx(...)} and recovers the coordinate with
\cpp{GetMomentumCoordFromFineIdx(...)} only when necessary. 

The bosonic bare vertices $\mathcal{B}_\X$ for each channel have names ending in \cpp{_bos}. They are specified by defining
\begin{lstlisting}[language=C++]
static double B_sc(int idx_W, int idx_K,
                                 int idx_w, int idx_m, 
                                 int idx_wp, int idx_mp);
\end{lstlisting}
and analogously for channels $\D$ and $\M$. Here \cpp{idx_W} and \cpp{idx_K}
are the bosonic transfer frequency and momentum indices, \cpp{idx_w} and
\cpp{idx_wp} are the fermionic relative-frequency indices, and \cpp{idx_m},
\cpp{idx_mp} are form-factor indices. The dependence on the form-factor indices is
essential: $\mathcal{B}_\X$ is already projected into the form-factor basis. Given the purely bosonic nature of $\mathcal{B}_\X$, this function returns a non-zero value only when \cpp{idx_m},\cpp{idx_mp} $\neq 0$.

The fermionic bare vertices $\mathcal{F}_\X$ follow an identical signature but
carry the suffix \cpp{_ferm}:
\begin{lstlisting}[language=C++]
static double F_sc(int idx_W, int idx_K,
                                  int idx_w, int idx_m,
                                  int idx_wp, int idx_mp);
\end{lstlisting}
and analogously for $\D$ and $\M$. Like the bosonic contributions, these functions are
projected into the form-factor basis and therefore depend on \cpp{idx_m} and
\cpp{idx_mp}. In models with nearest-neighbor density-density interactions, for
instance, the $\mathcal{F}_X$ functions are diagonal in form-factor space and
non-zero only for first-shell indices.

\paragraph{Double-counting terms.}
To construct the $\mathcal{B}$-irreducible vertex --- stored in the SBE state
and accessed via \cpp{state.B_irreducible_vertex_sc(...)}, \cpp{state.B_irreducible_vertex_d(...)},
\cpp{state.B_irreducible_vertex_m(...)} --- one needs the
double-counting contributions from the SBE decomposition ~\cite{AlEryani2025b}. These are provided by the model via
\cpp{vertex_DC_sc(...)}, \cpp{vertex_DC_d(...)}, and \cpp{vertex_DC_m(...)},
each with the same index signature as the bosonic channel objects. Since the
double-counting terms live in the same projected representation, they also carry
form-factor indices \cpp{idx_m} and \cpp{idx_mp}. For a model with a density-density bare interaction, the double-counting term is canonically trivial, namely minus two times the local part of the interaction; see also discussion in ~\cite{AlEryani2024}.

\paragraph{Local component of the vertex.}
The model class also provides the local component of the bare interaction through two
overloads of \cpp{vertex_local_part_bare(...)}. One overload takes the
channel-projected indices \cpp{(idx_W, idx_w, idx_m, idx_wp, idx_mp)}, and the
other takes the full set of four fermionic frequency--momentum--spin indices as
in \cpp{vertex_4pt_bare(...)}. These are used in postprocessing for fluctuation diagnostics. We briefly note that one may additionally define flowing bare vertices specific to the model itself. If available, these then specify how the bare vertex flows when the \cpp{FREE_U_FLOW} scheme is used. This has been used in the Holstein Anderson-impurity model to design a temperature flow, see Ref.~\cite{AlEryani2025b} as well as the files~\directory{include/models/anderson_impurity_holstein.h}, \directory{include/models/anderson_impurity_holstein.cpp} and \directory{src/models/anderson_impurity_holstein_flowing.cpp} for details.  

\subsubsection{Writing a new model}
Once the class header and its corresponding \directory{.cpp} file are ready, the
header should be included in \directory{include/models/concrete_available_models.h}.
The model can then be selected at compile time in \directory{config.mk} via the flag \cpp{-DTHE_MODEL_VAL=MyModel}, where \cpp{MyModel} is the model class name. Because every other part
of the code is templated on the model type, a correctly implemented model
immediately becomes available for all flow-equation types, frequency-grid settings,
output routines, and solver strategies without any further modifications.

\subsection{State and RHS Classes}
An instance of a state class stores the running many-body objects (primarily the self-energy $\Sigma$ and the SBE objects $w_\X$, $\lambda_\X$, $M_\X$) as a high dimensional vector, whereas an instance of an RHS class is a functor that computes either the state derivative for a flow equation or the next iterate in a self-consistency cycle. 

\subsubsection{States}
On the state side, \cpp{state_vector_frg_base_t<Model, OtherTypes...>} is the base class template. It stores the self-energy and the chemical-potential shift (determined such that the filling remains fixed), while the variadic template parameter \cpp{OtherTypes...} allows additional state components to be appended in a flexible way. The components of the state (self-energies or SBE objects) are represented by instances of a \cpp{gf} type, a lightweight wrapper around \cpp{boost::multiarray}. The \cpp{gf} container is based on a publicly available header-only implementation with convenience functions for symmetry reduction~\cite{wentzell_gf_container}, which we adopt with some modifications. Variants of this container have also been used in the \cpp{TRIQS} library~\cite{Parcollet_2015}, and a Julia implementation focused on Matsubara-frequency objects is available in the package \texttt{MatsubaraFunctions.jl}~\cite{Kiese_2024}.

The SBE state,
\cpp{state_frg_sbe_t<Model, OtherTypes...>}, builds on this base by adding
$w$, $\lambda$, $M$, and free-energy objects. The bosonized variant
extends it further by adding ``bosonic propagators'' \cpp{w_M} and ``fermion-boson'' vertices \cpp{lambda_M} from which the rest functions $M_\X$ can be reconstructed, see Refs.~\cite{Bonetti2022,Fraboulet2022}. This provides a natural extension strategy in which new state classes can reuse existing ones and simply extend them with new components. The chosen state type (determined by the selected flag combination) is then used for both the physical state and its derivative. States are initialized with
\cpp{state.init_bare()} which sets the components to the bare values, whereas state derivatives are
typically initialized with \cpp{init_zero()}. For interpolating schemes (such as in DMF2RG), the state can be initialized from reference DMFT input. For resumed calculations, the \cpp{FileIO} layer can restore the state from intermediate HDF5 files.

The commonly used state classes currently in use provide convenience methods that can be used for the RHS calculation or observables. For example, the state implements methods to construct SBE objects such as the full vertex and the irreducible vertices, and to compute inter-channel projections.
For the latter, these methods make direct use of precomputed TU projection matrices, which are implemented separately in the static class \cpp{TUProjections<Model>}.

If one wants to add a new state member, the idea then is conceptually simple. One overloads the state and supplements the variadic template parameter with new corresponding \cpp{gf} types. One overloads the initialization and optionally writes new accessor methods. 

\subsubsection{Right-hand sides}
Implementations of RHS classes follow a hierarchy: A base class \cpp{rhs_frg_base_t<Model, state_t>} provides general methods that are used throughout the code, such as the calculation of bubbles as convolutions of two propagators. Note that they are templated on the Model and the state classes. Further RHS classes inherit from this base class and implement the RHS calculations. An RHS class is primarily a functor, meaning that it defines the method
\begin{lstlisting}[language=C++]
void operator()(const state_t& old_state, state_t& dstate_over_dt, double t).
\end{lstlisting}
It implements an RHS that generically depends on a flow parameter, the state and the state derivative. The state-derivative \cpp{dstate_over_dt} is typically only needed in multiloop schemes, where the RHS also depends on derivatives of the state (e.g. already at the level of the Katanin substitution, $\dot{\Sigma}$ is needed). For example, \cpp{rhs_sbe_1lfrg_t} builds on the base class with one-loop SBE flow logic, and
\cpp{rhs_sbe_mfrg_t} extends it with multiloop-specific correction members, bookkeeping, and methods for computing higher-loop corrections. This inheritance chain keeps multiloop
complexity out of the one-loop core while still reusing all the shared machinery.

An important design detail is that all physics equations of the RHS are implemented separately from the rest in the \directory{rhs_*_eval.tpp} files. In this way, the physics equations are separated from the program control flow, and can be altered or investigated separately.

When an RHS functor is executed, the common pattern is first to compute all bubbles or bubble derivatives. Then individual calls are made to evaluate the RHS of each state variable. Here, OpenMP parallelization is used to evaluate the separate indices while exploiting symmetries (implemented in the \cpp{Symmetries*<Model>} classes) by calling
\cpp{init_batched(...)} or \cpp{init_batched_at_sampling_indices(...)}, then, if the log-frequency spacing of frequencies flag is enabled,
fill the remaining points via interpolators

\noindent
(\cpp{init_batched_at_interpolating_indices(...)}).

Implementing a new RHS amounts simply to adding another class to this hierarchy, following the same conventions as above: it should inherit from the appropriate base class. Additionally, the control flow should remain in the
main \cpp{rhs_*.tpp} file, while the physics equations should be kept in a dedicated \cpp{*_eval} file; symmetry-batched evaluation and the existing bubble/projection utilities should be reused, and the type alias \cpp{rhs_t} should be defined in \directory{start.h} when the corresponding preprocessor flag is enabled. Model information,
form-factor projection matrices, and symmetry structure remain reusable across all RHS variants.

\subsection{Observables and HDF5 Output}

The observer class serves two roles. During an fRG run, \cpp{observer_frg_t} is called at accepted
integration steps and checks whether any component of the vertex has diverged (signaling an instability). Second, it stores an instance of the \cpp{observables_t} class (templated on both the model and the state), which is responsible for calculating and storing observables, e.g., from the state. Outputting the observables is then delegated to the FileIO class,
which writes observables data to a HDF5 file. The FileIO, as its name suggests, is responsible for both writing and reading HDF5 files.

\subsubsection{Observable containers and update logic}
The common observable container is
\cpp{observables_frg_common_t<Model, state_t>} which contains basic observables such as the filling and the current scale. The class is inherited by
\cpp{observables_frg_sbe_t<Model, state_t>} which introduces SBE related observables. There are two types of observables contained in two containers: one returned by the static method \cpp{ObservablesListToTrack()}, which stores information about individual observables, and the other returned by \cpp{GroupObservablesListToTrack()}, whose entries describe groups of related observables that are written separately into HDF5 groups. At initialization, data containers are allocated and then populated each time the \cpp{update(...)} method is called. Additional observables calculated from an RHS class \cpp{rhs_postproc_sbe_t} include more heavy observables calculated from integral equations such as full susceptibility matrices and the constituents entering fluctuation diagnostics.

\subsubsection{Adding new observables}
Adding a new observable is done in three steps. First, it must be added to the tracking list in the \cpp{SetToTrackAllAvailableObservables()} method; this includes specifying the name of the observable, its type, dimension and name (to be used in the HDF5 output  path). Then the observable calculation must be implemented in the \cpp{update(...)}.

\subsection{Flow Schemes}

All flow schemes are implemented in the static class \cpp{fRGFlowScheme<Model>}, which is
templated on the same \cpp{Model}. Its role is limited to the scale dependence of
the propagators and related flow-dependent quantities.

\subsubsection{Choosing the flow}
The central call for ``choosing'' a flow is the following method called during startup:
\begin{lstlisting}[language=C++]
fRGFlowScheme<Model>::UseFlow(G_FlowSchemeName g_scheme, U_FlowSchemeName u_scheme);
\end{lstlisting}
called in \directory{start.cpp}. It selects two parts of the flow independently.
The \cpp{G_FlowSchemeName} choice determines how the propagators are cut off, in
particular the flowing Green's function \cpp{G}, its ``physically rescaled'' variant 
\cpp{G_latt}, and the single-scale propagator \cpp{S}. The
\cpp{U_FlowSchemeName} choice determines how the bare interaction is scaled during
the flow through functions such as \cpp{U_MultiplicativeCutoff(...)} and its
derivative. In this way, one may combine a propagator cutoff with an interaction
cutoff while keeping the same abstract flow parameter \cpp{t}.

Importantly, \cpp{UseFlow} simply registers the corresponding member functions so that the rest
of the code can always call the same interface. The solver and RHS therefore do
not need to know which specific flow scheme has been chosen, in line with the SoR principle.

\subsubsection{Flow-dependent propagators}
The flow-scheme dependent propagator
functions are called with a fermionic frequency index \cpp{idx_w}, a fine-grid
momentum index \cpp{idx_p} as well as the flow parameter \cpp{t}, the self-energy, and the
chemical-potential shift. At this stage, the flow-scheme code makes direct use of model information. In
particular, when the Green's function is constructed, the underlying dispersion
enters through the model, namely via \cpp{Model::E(...)}. Thus the flow scheme
controls how the cutoff enters the propagator, while the model still provides the dispersion.

\subsubsection{Adding a new cutoff scheme}
Adding a new cutoff scheme can be done almost exclusively within the \cpp{fRGFlowScheme<Model>} class. One introduces
the methods implementing the flowing propagator \cpp{G}, the
single-scale propagator \cpp{S}, as well as
\cpp{G_latt}. Optionally, one may add the asymptotic tail functions for the bubbles, which improve the accuracy of the Matsubara sums in the evaluation of the RHS. Next, one adds a new case to a switch statement found in the \cpp{UseFlow} method so that these functions are selected when
the corresponding enum value is chosen. New \cpp{U}-flow schemes can be implemented similarly.

\section{Conclusion and outlook}

We have presented \texttt{BosonFlow}, a C++ implementation of the functional renormalization group (fRG) within the single-boson exchange (SBE) formulation, providing a unified and flexible platform for quantitatively accurate calculations of dynamic vertex structures and self-energies in lattice fermion systems and for the development of new approximations, which yield to the physical transparency in the vertex constituents presented in the SBE decomposition. By combining the truncated-unity treatment of momentum dependence with the natural classification of frequency asymptotic classes present in the single-boson exchange decomposition of the vertex, the codebase addresses a critical computational bottleneck: it enables the simultaneous resolution of both momentum and frequency dependencies that have traditionally forced a costly trade-off in fRG applications. Our implementation emphasizes separation of concerns, with distinct modules for model specification, flow-scheme selection, state management, and observable calculation. This architectural principle ensures that extensions to new models, approximation schemes, or solution methods require minimal disruption to the existing codebase.

Looking forward, the  \texttt{\texttt{BosonFlow}} provides a robust platform for continued methodological development. Promising directions include the utilization of efficient frequency bases such as intermediate representation (IR)~\cite{shinaoka2017compressing,chikano2018performance} and the discrete Lehmann representation (DLR)~\cite{kaye2022discrete,kaye2022libdlr}, treatment of orbital dependence, and exploration and development of more efficient approximations to the SBE rest functions. Furthermore, development of new flow schemes, such as a finite-temperature adapted sharp cutoffs~\cite{Enss_2005}, and interpolation flow schemes that take into account correlated starting points from both the bare propagator or the bare interaction, are in principle straightforward, and worth investigating. In the future, we wish to address many of these problems.

The \texttt{BosonFlow} codebase is released as open source to serve both as a reference implementation for the fRG and SBE methods and as a foundation for future research in quantitative studies of strongly correlated quantum systems.

\section*{Acknowledgement}
The authors thank F.~Domizio, H.~Essl, M.~Gievers, F.~Giuseppe, A.~Kauch, N.~Gneist, M. Krämer, L.~Philoxene, J.~Profe, N.~Ritz, D.~Villardi, M.~Wallerberger, and A.~Walther for discussions, collaborations on related projects, and useful comments during the development of \texttt{BosonFlow}.

We acknowledge S. Heinzelmann, C. Hille, A. Tagliavini, and N. Wentzell for their contributions to an earlier functional renormalization group codebase in the past in the group of S. Andergassen, which informed the design of the current implementation.

A.~A.-E. thanks M.~Scherer for discussions, collaboration on related projects, and for providing the environment to conduct this research.

We thank S.~Andergassen for discussions and continuous feedback throughout the development of \texttt{BosonFlow} through various application projects.

\paragraph{Funding information}
A.A.-E.\ acknowledges funding from the Deutsche Forschungsgemeinschaft (DFG, German Research Foundation) under Project No.~277146847 (SFB 1238, project C02). K.F.\ acknowledges financial support from the Austrian Science Fund (FWF) 10.55776/I6946 and from the DFG within the research unit FOR 5413/1 (Grant No. 465199066).




\bibliography{bibliography}

\nolinenumbers

\end{document}